\documentstyle[graphics]{mn}

\def\ni{\noindent}                                       %No indent%
\def\etal{et\thinspace al.\ }                               %et al.%
\def\gapprox{_>\atop{^\sim}}       %Greater than over approximately%
\def\lapprox{_<\atop{^\sim}}          %Less than over approximately%
 
\newcommand{\ET}[1]{\times 10^{#1}}             %times power of ten%

\newcommand{\ov}[1]{\overline{#1}}                        %overline%

\newbox\grsign \setbox\grsign=\hbox{$>$} \newdimen\grdimen \grdimen=\ht\grsign
\newbox\simlessbox \newbox\simgreatbox
\setbox\simgreatbox=\hbox{\raise.5ex\hbox{$>$}\llap
     {\lower.5ex\hbox{$\sim$}}}\ht1=\grdimen\dp1=0pt
\setbox\simlessbox=\hbox{\raise.5ex\hbox{$<$}\llap
     {\lower.5ex\hbox{$\sim$}}}\ht2=\grdimen\dp2=0pt
\def\simgreat{\mathrel{\copy\simgreatbox}}

\title[Empirical Population Synthesis: Sampling methods and tests]
       {A probabilistic formulation for Empirical Population
       Synthesis: Sampling methods and tests}

\author[R. Cid Fernandes, L. Sodr\'e., H. Schmitt \& J. Le\~ao]
	{Roberto Cid Fernandes$^{1,2}$\thanks{e-mail: cid@fsc.ufsc.br}\thanks{Gemini fellow},
	Laerte Sodr\'e Jr.$^{3}$\thanks{e-mail: laerte@iagusp.usp.br}, 
	Henrique R. Schmitt$^{4}$\thanks{e-mail: schmitt@stsci.edu},\cr 
	Jo\~ao R. S. Le\~ao $^{2}$\thanks{e-mail: joao@fsc.ufsc.br} \\
  $^1$Depart.\ of Physics \& Astronomy, Johns Hopkins University, 
3400 N. Charles St., Baltimore, MD 21218, USA\\
  $^2$Depto.\ de F\'{\i}sica - CFM - Universidade Federal de Santa Catarina, 
CP 476, Campus Universit\'ario, Trindade,\\
88040-900, Florian\'opolis, SC, Brazil\\
  $^3$Depto.\ de Astronomia, Instituto Astron\^omico e Geof\'\i sico -
  USP, Av. Miguel Stefano 4200, 04301-904 S\~ao Paulo, Brazil \\
  $^4$ Space Telescope Science Institute, 3700 San Martin Drive, Baltimore,
  MD 21218, USA}

\begin{document}

\maketitle

%AAAAAAAAAAAAAAAAAAAAAAAAAAAAAAAAAAAAAAAAAAAAAAAAAAAAAAAAAAAAAAAAAAA
\begin{abstract} 
We revisit the classical problem of synthesizing spectral properties
of a galaxy using a base of star clusters, approaching it from a
probabilistic perspective.  The problem consists of estimating the
population vector ${\bf x}$, composed by the contributions of
$n_\star$ different base elements to the integrated spectrum of a
galaxy, and the extinction $A_V$, given a set of absorption line
equivalent widths and continuum colors. The formalism is applied to
the base of 12 elements defined by Schmidt \etal (1991) as
corresponding to the principal components of the original base
employed by Bica (1988), and subsequently used in several studies of
the stellar populations of galaxies. The exploration of the 13-D
parameter space is carried out with a Markov chain Monte Carlo
sampling scheme based on the Metropolis algorithm. This produces a
smoother and more efficient mapping of the $P({\bf x},A_V)$
probability distribution than the traditionally employed uniform-grid
sampling. 

This new version of Empirical Population Synthesis is used to
investigate the ability to recover the detailed history of star
formation and chemical evolution using this spectral base. This is
studied as a function of (1) the magnitude of the measurement errors
and (2) the set of observables used in the synthesis. Extensive
simulations with test galaxies are used for this purpose. Emphasis is
put on the comparison of input parameters and the {\it mean} ${\bf x}$
and $A_V$ associated with the $P({\bf x},A_V)$ distribution. It is
found that only for extremely low errors ($S/N > 300$ at 5870 \AA) all
12 base proportions can be accurately recovered, though the
observables are recovered very precisely for any $S/N$.  Furthermore,
the individual $\ov{\bf x}$ components are {\it biased} in the sense
that components which carry a large fraction of the light tend to
share their contribution preferably among components of same age.
Old, metal poor components can also be confused with younger, metal
rich components due to the age $\times$ metallicity degeneracy.  These
compensation effects are linked to noise-induced linear dependences in
the base, which very effectively redistribute the likelihood in ${\bf
x}$-space. The {\it age} distribution, however, can be satisfactorily
recovered for realistic $S/N$ ($\sim 30$). We also find that
synthesizing equivalent widths and colors produces better focused
results that those obtained synthesizing only equivalent widths,
despite the inclusion of the extinction as an extra parameter.

\end{abstract}

%AAAAAAAAAAAAAAAAAAAAAAAAAAAAAAAAAAAAAAAAAAAAAAAAAAAAAAAAAAAAAAAAAAA

\section{Introduction}

\label{sec:Introduction}

The stellar populations of galaxies carry a record of their
star-forming and chemical histories, from the epoch of formation to
the present. Their study thus provide a powerful tool to explore the
physics of galaxy formation and evolution.  Despite the detailed
understanding of stellar evolution and of the properties of single
stellar populations such as star clusters achieved over the past
century, uncovering the history of galaxies through the study of their
integrated light has proven to be a difficult problem (Pickles 1985;
Worthey 1994 and references therein).

Two global approaches have been developed to tackle this issue.  The
first of these is Evolutionary Population Synthesis, which performs
{\it ab initio} calculations of the spectral evolution of galaxies on
the basis of stellar evolution theory, stellar spectral libraries, and
prescriptions for the initial mass function, star formation rate and
chemical evolution (Tinsley 1972; Larson \& Tinsley 1974, 1978;
Guiderdoni \& Rocca-Volmerange 1987; Fioc \& Rocca-Volmerange 1997;
Charlot \& Bruzual 1991; Bruzual \& Charlot 1993; Leitherer \etal
1996, 1999 and references therein).  The other approach is Empirical
Population Synthesis (EPS hereafter), also known as `Stellar
Population Synthesis with a Database', which uses observed properties
of stars or star clusters as a {\it base} on which to decompose the
population mix in galaxy spectra (Spinrad \& Taylor 1971; O'Connell
1976; Faber 1972; Pickles 1985; Bica 1988; Pelat 1997, 1998; Boisson
\etal 2000). Of particular interest is the EPS method of synthesizing
the spectra of galaxies using a base of star clusters.  Perhaps the
most appealing property of this method is the fact that the results do
not hinge on assumptions about stellar evolution tracks nor initial
mass functions, as these are `mother nature given' in the empirically
built base spectra.  This technique has been pioneered by the work of
Bica (1988, hereafter B88), where the stellar content of a large
sample of galaxies was synthesized using the spectral base of star
clusters described in Bica \& Alloin (1986a, 1986b, 1987). Since then,
several studies made use of this technique (Schmidt, Bica \& Dottori
1989; Bica, Alloin \& Schmidt 1990; Jablonka, Alloin \& Bica 1990;
Schmidt, Bica \& Alloin 1990; Jablonka \& Alloin 1995; De Mello \etal
1995; Bonatto \etal 1996, 1998, 1999, 2000; Schmitt, Storchi-Bergmann
\& Cid Fernandes 1999; Kong \& Cheng 1999; Raimann \etal 2000).

In its original version (B88), this method made use of a large base of
$n_\star = 35$ elements, combined in different proportions ($x_i$; $i
= 1 \ldots 35$) to synthesize a number of equivalent widths ($W$) of
conspicuous absorption features. An uniform sweep of the parameter
space along paths consistent with simple predefined chemical evolution
scenarios was performed, and all combinations which produced $W$'s
within 10\% of the observed values were considered `solutions'. The
final population vector ${\bf x}$ was taken to be the {\it arithmetic
mean} of all such sampled solutions.

One of the main concerns raised by B88 method is that of the {\it
uniqueness} of the solution, as first discussed by Schmidt \etal
(1991, hereafter S91). With a 35-D population vector and at most 9
$W$'s to be synthesized, one has a highly degenerate algebraic
system. As shown by S91, however, the original base was highly
redundant, with several linearly dependent elements, and others which
were practically so, as deduced by a principal component analysis. S91
then proposed the use of a reduced base of 12 elements. S91 also
criticized B88's uniform sampling of the ${\bf x}$ space (a `discrete
combination procedure' in their terms) and the use of a `mean
solution'.  Instead, they developed a constrained minimization
procedure, which, as shown by their simulations with test galaxies
built out of the base, is able to retrieve with good accuracy the
correct population vector. Another difference of the S91 method
compared to B88 is the possibility to search for solutions along the
entire age $\times$ metallicity plane. B88 assumed that the chemical
enrichment of the galaxy occurred during its formation ($\sim10$ Gyr
ago), with all stars younger than that having the same $Z$ as the
oldest most metal rich component. Although this may be a reasonable
approximation for the nuclear region of isolated galaxies, it does not
allow for the effects of mergers, inflows or outflows, and is not
applicable to high redshift sources, where the spectra usually
integrate over the entire galaxy, including regions with different
star formation and chemical histories (Jablonka \etal 1990; Jablonka
\& Alloin 1995).

The next improvement in this technique was the introduction of
continuum colors in the synthesis. This was done by Schmitt, Bica \&
Pastoriza (1996), who, as B88, use a discrete combination procedure
search for acceptable solutions, sweeping the parameter space along
chemical enrichment paths. This version of EPS has been used in
several subsequent studies (Bonatto \etal 1996, 1998, 1999, 2000; Kong
\& Cheng 1999). Another step was taken by Bonatto, Bica \& Alloin
(1995), who extended the star cluster base to the space ultraviolet
(1200--3200 \AA). They too use an arithmetic average to represent the
solution of the synthesis.

In two recent papers, Pelat revisited the EPS problem of synthesizing
$W$'s using an elegant formalism based on convex algebra concepts.  In
Pelat (1998) he shows how the algebraic solution of the EPS problem
can be narrowed down to a sub-space of ${\bf x}$ delimited by a set of
`extreme-solutions', any convex combination of which yields an exact
solution in the underdetermined cases (more parameters than
constraints). His results shed new light into the issue of algebraic
degeneracy, which has always haunted EPS due to the widespread belief
that with more parameters than observables one is bound to fit the
data in one way or another. He proposes a quick test of whether the
$W$'s of a galaxy are synthesizable by computing the region in
$W$-space spanned by all physical combinations of the base elements.
We verified that in the case of the 12 elements base used by S91 this
region is narrow (Le\~ao 2001), such that small measurement errors are
enough to place objects outside this `synthetic domain', hence
preventing the existence of mathematically exact solutions. In real
applications, one therefore expects the degeneracy of EPS to be more
of a statistical nature than algebraic. In these cases, as well as in
overdetermined problems, Pelat (1997) demonstrates that the best model
is to be found along the boundary of the synthetic domain by
minimizing an appropriately defined $\chi^2$-like distance.
Uncertainties in the population vector can also be readily computed,
as shown by Moultaka \& Pelat (2000).

Boisson \etal (2000) recently applied Pelat's method to a sample of 12
active galaxies from Serote Roos \etal (1998). They synthesize $n_W =
47$ absorption lines with a base of $n_\star = 30$ stars from the
stellar spectral libraries of Serote Roos, Boisson \& Joly (1996) and
Silva \& Cornell (1992). Their results nicely illustrate the power of
this new EPS technique, which will certainly be extended to larger
samples and other classes of galaxies in future studies.  The B88
method and its variants, on the other hand, already have a large body
of literature associated to it. Nevertheless, this popular and
attractive EPS technique, while intuitively acceptable, has neither
been adequately tested nor cast onto a sound mathematical formalism
which supports its application. Furthermore, it gives no measure of
the uncertainties or potential biases on the population vector, thus
giving no simple way to prevent overinterpretations of the resulting
population of the age $\times$ metallicity plane. These issues are the
focus of the present paper.

Our main goals are to:	

\begin{itemize}

\item[(i)] Elaborate a mathematically consistent formulation of the
EPS problem in the context of probability theory.

\item[(ii)] Develop and test a sampling method to explore the
parameter space in EPS problems. In particular, we aim to improve upon
the Schmitt \etal (1996) technique of synthesizing both equivalent
widths and colors, but within the context of a probabilistic
formulation.

\item[(iii)] Use the method to investigate the effects of measurement
errors and of the specific set of observables used upon the results of
the synthesis with the reduced B88 star cluster base described in S91.
In particular, we wish to evaluate whether these factors introduce
biases in the population mix inferred from this popular synthesis
technique and at what level of detail can its results be trusted.

\end{itemize}

This paper is organized as follows: Item {\it (i)} of the above list
is addressed in section~\ref{sec:formalism}. In section
\ref{sec:Test_Metropolis} we deal with point {\it (ii)}. This is done
by investigating the applicability of a Metropolis algorithm to sample
the parameters probability distribution in EPS, and testing the method
with simulations based on Bica's spectral base. In section
\ref{sec:Test_of_Bicas_Base} we address item {\it (iii)} by means of
extensive simulations which explore the ability to recover the
detailed star-formation and chemical histories of galaxies with this
base. Finally, section \ref{sec:conclusions} summarizes our main
results.

\section{Formalism}

\label{sec:formalism}

\subsection{Basic equations}

EPS studies seek to find combinations of a spectral base which
reproduce a given set of measured observables, often taken as the
equivalent widths $W_j$ of $n_W$ conspicuous absorption features. The
base consists of $n_\star$ elements representing well defined simple
stellar populations such as star-clusters, with equivalent widths
$W^\star_{ij}$ and corresponding continuum fluxes over the lines
$F^\star_{ij}$ ($j = 1, \ldots n_W$, $i = 1, \ldots n_\star$)
normalized at a reference wavelength. Denoting by $x_i$ the fractional
contribution of the $i$-th base element to the total flux at the
reference wavelength, one obtains a system of $n_W$ equations

\begin{equation}
\label{eq:W_syn}
W_j = W_j({\bf x}) =
  \frac{\sum_{i=1}^{n_\star} W^\star_{ij} F^\star_{ij} x_i}{
        \sum_{i=1}^{n_\star} F^\star_{ij} x_i}; ~~~~j = 1,...,n_W
\end{equation}

\ni for the synthetic $W$'s (e.g., Joly 1974). This set of constraints
can be augmented by synthesizing $n_C$ observed continuum fluxes $C_k$
($k = 1, \ldots n_C$), provided allowance is made for reddening, here
parametrized by the V-band extinction $A_V$. Assuming all $n_\star$
populations are equally reddened, one obtains

\begin{equation}
\label{eq:C_syn}
C_k = 
  C_k({\bf x},A_V) =
  g_k(A_V) \sum_{i=1}^{n_\star} C^\star_{ik} x_i; ~~~~k = 1,...,n_C  
\end{equation}

\ni where a distinction is made between $C^\star$ and $F^\star$, since
the $n_C$ fluxes to be synthesized need not correspond to the
wavelengths of the $n_W$ absorption lines. We hereafter refer to the
$C_k$'s as continuum {\it colors}, as they are in fact ratios of
continuum fluxes with respect to the continuum at the normalization
wavelength.  The function $g_k(A_V)$ reddens the normalized color at
wavelength $\lambda_k$ by $A_{\lambda_k}$ according to a specified
reddening law.

Finally, physical solutions must further satisfy the normalization and
positivity constraints:

\begin{equation}
\label{eq:normalization}
\sum_{i=1}^{n_\star} x_i = 1;
~~x_i \ge 0~{\rm for}~ i = 1,...,n_\star,~{\rm and}~~A_V \ge 0.
\end{equation}

The normalization condition effectively reduces one degree of freedom.
When modeling colors, however, one has to introduce $A_V$\ as a
further parameter, so that the number of parameters still is
$n_\star$.

\subsection{Probabilistic formulation}

The data ${\cal D}$\ to be modeled are thus composed of a set of
$n_{obs} = n_W + n_C$ observables. The measurement errors in each
observable, collectively denoted by ${\bf \sigma}$, are assumed to be
known.  Given these, the problem of EPS is to estimate the population
vector ${\bf x}$ and the extinction $A_V$\ that `best' represents the
data according to a well defined probabilistic model. It is equally
important to estimate the uncertainties in the parameters, as these
prevent over-interpreting the resulting mixture of stellar populations
at a level of detail not warranted by the data or by intrinsic
limitations of the base.

The probability of a solution $({\bf x}, A_V)$ given the data ${\cal
D}$ and the errors ${\bf \sigma}$, is given by Bayes theorem:

\begin{equation}
\label{eq:Bayes}
P({\bf x}, A_V | {\cal D}, {\bf \sigma}, {\cal H}) = 
  \frac{P({\cal D} | {\bf x}, A_V,{\cal H})
  P({\bf x}, A_V | {\bf \sigma} , {\cal H})}
  {P({\cal D} | {\cal H})}.
\end{equation}

${\cal H}$ summarizes the set of assumptions on which the inference is
to be made. They include: the mapping between parameters and
observables is given by eqs.~(\ref{eq:W_syn}) and (\ref{eq:C_syn});
${\bf x}$ and $A_V$ must satisfy the constraints expressed in
eq.~(\ref{eq:normalization}); the stellar population in the target
galaxy is well represented by the base elements; the observational
errors are Gaussian; the reddening law is known.

The likelihood $P({\cal D} | {\bf x}, A_V, {\bf \sigma}, {\cal H})$ is
a measure of how good (or bad) is the fitting to the data for model
parameters ${\bf x}$ and $A_V$. Under the hypothesis of Gaussian
errors, the probability of the data given the parameters is

\begin{equation}
\label{eq:likelihood}
P({\cal D} | {\bf x}, A_V, {\bf \sigma}, {\cal H}) 
  \propto e^{-{\cal E}},  
\end{equation}

\ni with ${\cal E}$ defined as half the value of $\chi^2$:

% This version screws up in the pre-print format ...
%\begin{equation}
%\label{eq:energy}
%{\cal E}({\bf x}, A_V) = 
%  \frac{1}{2} \chi^2({\bf x},A_V) =
%  \frac{1}{2}  \sum_{j=1}^{n_W}
%    \left( \frac{ W_j^{obs} - W_j({\bf x}) }{\sigma(W_j)} \right)^2 +
%  \frac{1}{2}  \sum_{k=1}^{n_C}
%    \left( \frac{ C_k^{obs} - C_k({\bf x},A_V) }{\sigma(C_k)} \right)^2 
%\end{equation}

\begin{eqnarray}
\label{eq:energy}
{\cal E}({\bf x}, A_V) 
  & = &
    \frac{1}{2} \chi^2({\bf x},A_V) 
    =
    \frac{1}{2}  \sum_{j=1}^{n_W}
    \left( \frac{ W_j^{obs} - W_j({\bf x}) }{\sigma(W_j)} \right)^2 
    \\ \nonumber
  & &
    + \frac{1}{2}  \sum_{k=1}^{n_C}
    \left( \frac{ C_k^{obs} - C_k({\bf x},A_V) }{\sigma(C_k)} \right)^2 
\end{eqnarray}

\ni The likelihood, as the total $\chi^2$, separates into a $W$ and a
color related term, of which only the latter depends on $A_V$.

The normalizing constant $P({\cal D} | {\cal H})$ in Bayes theorem
(the `evidence of ${\cal H}$') is irrelevant to the level of
inference discussed here, i.e., the estimation of ${\bf x}$ and $A_V$.
It can be used, for instance, to compare different spectral bases.

$P({\bf x}, A_V | {\cal H})$ is the joint {\it a priori} probability
distribution of ${\bf x}$ and $A_V$, and states what values the model
parameters might plausibly take. For instance, physically acceptable
population fractions should have priors that are zero in regions of
the parameter space where the constraints of positivity and unit sum
are not satisfied. A prior that is uniform on the parameters and
includes these constraints is a {\it non-informative prior}, because
it does not impose any constraints on the solutions besides those
expressed by eq.~(\ref{eq:normalization}). This is the prior that will
be used in this work.

For a non-informative prior the posterior probability $P({\bf x}, A_V
| {\cal D}, {\bf \sigma}, {\cal H})$ is simply proportional to the
likelihood:

\begin{equation}
\label{eq:full_posterior}
P({\bf x}, A_V | {\cal D}, {\bf \sigma}, {\cal H}) 
  \propto e^{-{\cal E}({\bf x}, A_V)} 
\end{equation}

This expression contains the full solution of the EPS problem, as
embedded in it is not only the most probable model parameters but also
their full probability distributions. Furthermore, projected posterior
distributions for any of the $n_\star$ parameters can be obtained by
{\it marginalizing} eq.~(\ref{eq:full_posterior}) with respect to all
other parameters.  For instance, the probability density of proportion
$x_i$\ of the $i$-th base element is

%\begin{equation}
%\label{eq:post_x_i}
%P(x_i | {\cal D}, {\bf \sigma}, {\cal H}) = 
%    \int \ldots \int P({\bf x}, A_V | {\cal D}, {\bf \sigma}, {\cal H})
%    dx_1 \ldots dx_{i-1} dx_{i+1} \ldots dx_{n_\star} dA_V 
%\end{equation}

\begin{eqnarray}
\label{eq:post_x_i}
P(x_i | {\cal D}, {\bf \sigma}, {\cal H}) 
  & = & 
    \int \ldots \int P({\bf x}, A_V | {\cal D}, {\bf \sigma}, {\cal H})
    dx_1 \ldots \\ \nonumber
  & &
    dx_{i-1} dx_{i+1} \ldots dx_{n_\star} dA_V 
\end{eqnarray}

\ni and equivalently for $A_V$. Similarly, one can construct joint
posteriors for, say, the total proportion of all base components of
same age or $Z$, or for the mean age and $Z$ of the stellar
population. This would of course provide a coarser description of a
galaxy's stellar content than the individual posteriors $P(x_i | {\cal
D}, {\bf \sigma}, {\cal H})$, but that may be all that is possible
under some circumstances, such as when only a reduced set of
observables is available or when observational errors are large
(section \ref{sec:Test_of_Bicas_Base}).

It is easy to see how the B88 and Schmitt \etal (1996) synthesis
techniques fit into this general probabilistic formulation. By
synthesizing the observables within $\sim 10\%$ `error boxes', these
authors implicitly assumed a box-car likelihood function $P({\cal D} |
{\bf x}, A_V, {\bf \sigma}, {\cal H})$, whereas by performing
arithmetic means over all accepted $({\bf x},A_V)$ combinations in a
uniform grid they are implicitly sampling the corresponding posterior
probability distributions. Also, the {\it a priori} constraints on the
occupation of the age $\times$ $Z$ plane imposed by these authors (but
relaxed in subsequent works) simply reflect their use of an
informative prior.  This gives some formal justification for their
heuristically designed EPS method.  It is therefore reasonable to
expect that an implementation of EPS based on the formalism presented
here should give results roughly compatible with those previously
obtained. An uniform sweep of the (${\bf x},A_V$) space is however a
very inefficient sampler for (\ref{eq:post_x_i}), so some improvement
is needed there. This is discussed next.

\section{Sampling Method and Tests}

\label{sec:Test_Metropolis}

Despite its formal merits, at the computational level our
probabilistic approach faces the same basic difficulty as previous EPS
codes, namely, the high dimensionality of the problem. For spectral
bases with astrophysically interesting resolution in age and
metallicity the number of elements $n_\star$ quickly becomes large
enough to render the exploration of the parameter space a non-trivial
task. Before discussing methods to sample the (${\bf x},A_V$) space,
we present the spectral base used in this work
(\S\ref{sec:Bicas_base}) and describe how we deal with the
uncertainties in the observables (\S\ref{sec:Error_recipe}).

\subsection{The spectral base and observables}

\label{sec:Bicas_base}

%***TAB***TAB***TAB***TAB***TAB***TAB***TAB***TAB***TAB***TAB***TAB
\begin{table*}
\begin{centering}
\begin{tabular}{cccccc}
\multicolumn{6}{c}{Base Elements Used}\\ \hline
HII & 10 Myr & 100 Myr & 1 Gyr & 10 Gyr & log(Z/Z$_{\odot}$)\\ \hline
    &   10   &   8     &  5    &   1    &  0.6\\
12  &   11   &   9     &  6    &   2    &  0.0\\
    &        &         &  7    &   3    & -1.0\\
    &        &         &       &   4    & -2.0\\ \hline
\end{tabular}
\end{centering}
\caption{Ages, metallicities and numbering convention for the star
clusters in the base.}
\label{tab:Base}
\end{table*}
%***TAB***TAB***TAB***TAB***TAB***TAB***TAB***TAB***TAB***TAB***TAB

The base used for the synthesis here is that of S91. It contains
$n_\star = 12$ population groups, spanning five age bins---10 Gyr
(which actually represent globular cluster-like populations), 1 Gyr,
100 Myr, 10 Myr and HII (corresponding to current star formation, and
represented by a pure $F_\lambda \propto \lambda^{-2}$ continuum based
on the spectrum of 30 Dor)---and four metallicities---0.01, 0.1, 1 and
4 Z$_\odot$ (Table 1). The observables in this base comprise the $n_W
= 9$\ equivalent widths of the absorption lines CaII~K~$\lambda$3933,
CN~$\lambda$4200, G~band~$\lambda$4301, MgI~$\lambda$5175,
CaII~$\lambda$8543, CaII~$\lambda$8662, H$\delta$, H$\gamma$ and
H$\beta$, as well as $n_C = 7$\ continuum fluxes at selected pivot
wavelengths: 3290, 3660, 4020, 4510, 6630, 7520 and 8700 \AA, all
normalized to 5870 \AA.  According to Bica \& Alloin (1986a, 1987) and
Bica, Alloin \& Schmitt (1994), the equivalent width windows and
continuum points were defined based on very high signal to noise
spectra of galaxies, with the express goal of using them to synthesize
the stellar population of galaxies. These spectra were obtained by
creating two average spectra, of blue and red galaxies, where the
stellar population features can be clearly traced. The base values for
these quantities have been previously published in Bica \& Alloin
(1986b, 1987) and Schmitt \etal (1996), and some were measured from
data in Bica \etal (1994). These are recompiled in Table 2 , as they
bear a direct impact on the analysis below.  There are thus a maximum
of $9 + 7 = 16$ observables to model with $n_\star = 12$ parameters:
$n_\star - 1$ population fractions plus $A_V$.  However, in several of
the tests presented below we shall make use of a reduced subset of
observables, as observational data sets seldom cover the whole
spectral range spanned by this base.

According to Bica \& Alloin (1986a, 1987) and Bica, Alloin \& Schmitt
(1994), since the ultimate goal of the continuum points and $W$'s is
to synthesize the stellar population of galaxies, the $W$
windows and continuum points were defined based on very high signal to
noise spectra of galaxies. These spectra were obtained by
creating two average spectra, of blue and red galaxies, where the
stellar population featuress can be identified.

The reddening of the colors is modeled with the extinction law
described in Cardelli, Clayton \& Mathis (1989, with $R_V = 3.1$).

%***TAB***TAB***TAB***TAB***TAB***TAB***TAB***TAB***TAB***TAB***TAB
\begin{table*}
\begin{centering}
\begin{tabular}{cccccccccc}
\multicolumn{10}{c}{Equivalent Widths (\AA)}\\ \hline
\#  & K & CN & G & MgI & CaT$_1$ & CaT$_2$ & 
 H$\delta$ & H$\gamma$ & H$\beta$  \\ \hline
 1 & 21.1 & 17.5 & 11.1 & 10.1 & 6.8 & 6.0 &  4.4 & 4.9 & 3.5\\
 2 & 17.3 & 12.0 &  9.3 &  7.4 & 5.5 & 5.0 &  4.4 & 4.9 & 3.5\\
 3 & 11.0 &  4.5 &  5.9 &  3.8 & 3.4 & 3.3 &  4.4 & 4.9 & 3.5\\
 4 &  4.7 &  0.5 &  2.5 &  0.7 & 1.2 & 1.6 &  4.4 & 4.9 & 3.5\\
 5 & 17.3 & 13.9 &  9.0 &  9.1 & 6.8 & 6.0 &  9.7 & 7.7 & 7.5\\
 6 & 14.0 &  9.6 &  7.4 &  6.2 & 5.5 & 5.0 &  9.7 & 7.7 & 7.5\\ 
 7 &  8.9 &  3.6 &  4.6 &  3.2 & 3.4 & 3.3 &  9.7 & 7.7 & 7.5\\   
 8 &  4.5 &  3.0 &  1.5 &  3.2 & 6.8 & 6.0 & 10.5 & 9.9 & 7.9\\   
 9 &  3.8 &  2.2 &  1.2 &  2.5 & 5.5 & 5.0 & 10.5 & 9.9 & 7.9\\   
10 &  2.6 &  1.4 &  0.3 &  2.5 & 8.2 & 6.9 &  4.5 & 3.5 & 3.9\\   
11 &  2.2 &  1.1 &  0.3 &  2.0 & 6.6 & 5.8 &  4.5 & 3.5 & 3.9\\   
12 &  0.0 &  0.0 &  0.0 &  0.0 & 0.0 & 0.0 &  0.0 & 0.0 & 0.0\\ \hline
\multicolumn{10}{c}{Continuum over the lines}\\ \hline
 1 & 0.34 & 0.48 & 0.55 & 0.87 & 1.03 & 1.04 & 0.44 & 0.58 & 0.79\\   
 2 & 0.48 & 0.59 & 0.65 & 0.90 & 0.96 & 0.96 & 0.56 & 0.67 & 0.83\\   
 3 & 0.72 & 0.78 & 0.81 & 0.95 & 0.84 & 0.84 & 0.76 & 0.82 & 0.91\\   
 4 & 0.94 & 0.96 & 0.97 & 1.01 & 0.72 & 0.71 & 0.96 & 0.97 & 1.00\\  
 5 & 1.03 & 1.05 & 1.06 & 1.04 & 0.70 & 0.70 & 1.05 & 1.06 & 1.06\\   
 6 & 1.03 & 1.05 & 1.06 & 1.04 & 0.70 & 0.70 & 1.05 & 1.06 & 1.06\\   
 7 & 1.03 & 1.05 & 1.06 & 1.04 & 0.70 & 0.70 & 1.05 & 1.06 & 1.06\\   
 8 & 1.89 & 1.75 & 1.68 & 1.23 & 0.59 & 0.58 & 1.79 & 1.65 & 1.38\\   
 9 & 1.89 & 1.75 & 1.68 & 1.23 & 0.59 & 0.58 & 1.79 & 1.65 & 1.38\\   
10 & 1.55 & 1.34 & 1.27 & 1.06 & 0.95 & 0.94 & 1.40 & 1.25 & 1.08\\ 
11 & 1.55 & 1.34 & 1.27 & 1.06 & 0.95 & 0.94 & 1.40 & 1.25 & 1.08\\ 
12 & 2.34 & 2.03 & 1.93 & 1.37 & 0.40 & 0.38 & 2.11 & 1.90 & 1.56\\ \hline
\end{tabular}
\begin{tabular}{cccccccc}
\multicolumn{8}{c}{Continuum Colors}\\ \hline
\# & 3290 & 3600 & 4020 & 4510 & 6630 & 7520 & 8700\\ \hline
 1 & 0.17 & 0.27 & 0.39 & 0.71 & 1.01 & 1.01 & 1.04\\   
 2 & 0.27 & 0.37 & 0.52 & 0.77 & 0.99 & 0.96 & 0.96\\   
 3 & 0.42 & 0.52 & 0.74 & 0.88 & 0.96 & 0.89 & 0.84\\   
 4 & 0.57 & 0.67 & 0.95 & 0.99 & 0.92 & 0.81 & 0.71\\   
 5 & 0.71 & 0.65 & 1.04 & 1.08 & 0.92 & 0.81 & 0.70\\   
 6 & 0.71 & 0.65 & 1.04 & 1.08 & 0.92 & 0.81 & 0.70\\   
 7 & 0.71 & 0.65 & 1.04 & 1.08 & 0.92 & 0.81 & 0.70\\   
 8 & 1.10 & 0.94 & 1.84 & 1.52 & 0.82 & 0.68 & 0.58\\   
 9 & 1.10 & 0.94 & 1.84 & 1.52 & 0.82 & 0.68 & 0.58\\   
10 & 2.10 & 1.52 & 1.45 & 1.11 & 0.95 & 0.99 & 0.94\\   
11 & 2.10 & 1.52 & 1.45 & 1.11 & 0.95 & 0.99 & 0.94\\   
12 & 3.78 & 2.56 & 2.20 & 1.73 & 0.74 & 0.56 & 0.37\\ \hline
\end{tabular}
\end{centering}
\caption{Observables in the base, which make the matrices
$W^\star_{ij}$, $F^\star_{ij}$ and $C^\star_{ij}$.  All continuum
fluxes are normalized to the continuum at 5870 \AA.}
\label{tab:Base_Colors}
\end{table*}
%***TAB***TAB***TAB***TAB***TAB***TAB***TAB***TAB***TAB***TAB***TAB

\subsection{Errors in the observables}

\label{sec:Error_recipe}

The measurement errors in the observables play a key role in
determining the structure of the likelihood function.  Whereas such
errors are available when analyzing observed galaxy spectra, they have
to be postulated in the synthesis of test galaxies performed below.
An alternative and potentially more attractive approach to deal with
the errors is to marginalize eq.~(\ref{eq:post_x_i}) over ${\bf
\sigma}$. This yields more conservative estimates for the parameters,
insofar as they are independent of the actual observational errors. In
this paper we follow the traditional approach of treating ${\bf
\sigma}$ as a relevant constraint in the synthesis process.

In order to insure a realistic correspondence between the values of
the errors and the actual quality of the data we have adopted a
prescription for the $\sigma$'s based on our experience in dealing
with the measurement of $W$'s and $C$'s in galaxy spectra (Cid
Fernandes, Storchi-Bergmann \& Schmitt 1998).  Two sources of
uncertainty affect the measurements of $W$'s: (1) the noise within the
line window, and (2) the uncertainty in the positioning of the
pseudo-continuum fluxes $C_k$ at the pivot wavelengths, which affect
$W_j$ because they define the continuum over the line. The effect of
the first of these sources upon $\sigma(W_j)$ is straight-forwardly
obtained through standard propagation of errors by specifying the
noise level at $\lambda_j$, the spectral resolution $\delta \lambda$,
and the size $\Delta_j$ of the pre-defined window over which $W_j$ is
measured. In Bica \& Alloin (1986a) system the pseudo continuum is
defined interactively by visual inspection of the spectrum around the
pivot $\lambda$'s, which makes the estimation of $\sigma(C_k)$ non
straight-forward. In Cid Fernandes \etal (1998) we found these errors
could be roughly scaled from the signal-to-noise ratio measured in
nearby `line-free' regions, such that the $S/N$ in the pseudo
continuum is $(S/N)_{\rm PC} \sim \rho_{\rm PC} (S/N)_\lambda$, with
$\rho_{\rm PC} \sim 2$--3. This leads to the following expression for
the $\sigma(W_j)$'s:

\begin{equation}
\sigma^2(W_j) = 
  \Delta_j \delta_\lambda (S/N)_\lambda^{-2} +
  (\Delta_j - W_j)^2 \rho_{\rm PC}^{-2} (S/N)_\lambda^{-2}
\end{equation}

\ni where the two terms correspond to the two sources of uncertainty
discussed above. For simplicity, we further assume that the noise in
the observed spectrum is constant for all $\lambda$'s, which results
in $(S/N)_\lambda = C_\lambda (S/N)_{5870}$.  The errors in the
$C_k$'s thus become also constant:

\begin{equation}
\sigma(C_k) = \rho_{\rm PC}^{-1} (S/N)_{5870}^{-1}
\end{equation}

Besides yielding realistic values for the $\sigma$'s, this recipe has
the advantage of quantifying the quality of the data in terms of a
single quantity: The signal-to-noise ratio at the normalization
wavelength 5870 \AA. Of secondary importance are the spectral
resolution and $\rho_{\rm PC}$, which in the simulations below are
kept fixed at 5 \AA\ and 3 respectively.

In order to provide a quantitative notion of how a value of $S/N$
translates into $\sigma(W_j)$'s, we have computed these quantities for
the observables corresponding to spectral groups S1 and S7 of B88,
typical of the nuclei of early (red) and late (blue) type spiral
galaxies respectively. For S1 our error recipe yields $\sigma(W_{\rm
CaIIK}) = 36 {\rm \AA} / (S/N)_{5870}$ and $\sigma(W_{\rm MgI}) = 19
{\rm \AA} / (S/N)_{5870}$, so errors of $\sim 1$ \AA\ or less are
achieved with $(S/N)_{5870} \ge 30$. For S7 these lines have
$\sigma(W) \sim 15 {\rm \AA} / (S/N)_{5870}$.

\subsection{Exploration of the parameter space: Sampling methods}

\label{sec:Metropolis_method}

The exploration of the (${\bf x},A_V)$ space can be approached from
two alternative perspectives: (1) A {\it minimization} problem,
employing methods to search for the global $\chi^2$ minimum (maximum
likelihood). Minimization techniques were explored in S91 and Pelat
(1997) for the synthesis of $W$'s only.  (2) A {\it sampling} problem,
where one seeks to map out the posterior probability distribution
(eq.~\ref{eq:full_posterior}).

The methods discussed in this paper are {\it sampling} methods, which
do not explicitly search for a minimum. In the limit of small errors,
however, one expects the posterior maps to peak at the most likely
model. For large errors, on the other hand, the best model sampled may
be well off the true global minimum, but in this case the probability
distribution is so broad that the very meaning of `best model' is
questionable. In such cases it is arguably more important to estimate
plausible ranges for the parameters than to carry out a refined search
for the most likely values.

Our main motivation to explore the sampling approach is that most
applications of EPS with Bica's base to date focused on the estimation
of {\it mean} parameters, whose meaning cannot be directly ascertained
without estimating their uncertainty and/or biases.  A critical
re-evaluation of this method is thus crucial to establish to which
level of detail the results of this popular technique can be trusted.

\subsubsection{Uniform sampling}

In order to compute the individual posterior probabilities for each
parameter, we first note that eq.~(\ref{eq:post_x_i}) for the $x_i$
posterior is simply the {\it mean} probability $P({\bf x}, A_V | {\cal
D}, {\bf \sigma}, {\cal H})$ over the space spanned by all possible
values of $x_{j \ne i}$ and $A_V$ for a fixed $x_i$.  The simplest
method to compute such probabilities is to divide the (${\bf x},A_V$)
space into a {\it uniform} grid with $\Delta x$\ steps for the
population fractions and $\Delta A_V$ for the reddening parameter. One
then approximates eq.~(\ref{eq:post_x_i}) by the finite sum

\begin{eqnarray}
\label{eq:post_indiv_uniform_grid}
P(x_i | {\cal D}, {\bf \sigma}, {\cal H}) & \simeq & 
  \frac{ \sum_s e^{-{\cal E}({\bf x}_s,A_{V,s})} \delta(x_i - x_{i,s}) }
       { \sum_s e^{-{\cal E}({\bf x}_s,A_{V,s})} }
\end{eqnarray}

\ni where $s$ denotes a point (a `state') in the 13-D grid and the
$\delta(x_i - x_{i,s})$ term retains only those grid points where
population $i$ contributes with fraction $x_i$ of the light.

This discrete sweeping of the parameter space is easy to implement,
and was in fact the recipe followed by most works with Bica's base to
date (see section \ref{sec:Introduction}). A serious drawback of
uniform sampling is its computational price. The number of grid cells
is an extremely steep function of the resolution $\Delta x$. A coarse
$\Delta x = 10\%$ sampling yields $3.5\ET{5}$ ${\bf x}$-points,
increasing to $\sim 8\ET{7}$ for $\Delta x = 5\%$\ and more than
$4\ET{11}$\ for a $2\%$ resolution! To obtain the grid size one has to
further multiply these numbers by the number of points in the $A_V$
grid, whose limits have to be pre-defined.  Schmitt \etal (1999), for
instance, performed a synthesis study with $\Delta x = 5\%$ and
$\Delta A_V = 0.06$ for $A_V$ between 0 and 1.5, which yields
$2\times10^9$ grid points, perhaps the finest grid ever used in EPS
with this 12 elements base.  One obviously always aim for the best
resolution possible, but {\it a priori} estimates of what resolution
is necessary are not straight-forward. A `good' resolution should
produce probability distributions that are smooth over scales of
$\Delta x$ and $\Delta A_V$. The width of these distributions is of
course a function of the observational errors and of the set of
observables being synthesized.

\subsubsection{Metropolis sampling}

A further drawback of uniform sampling is that the algorithm spends
most of the time in regions of negligible probability, which
contribute little to the integral in equation~(\ref{eq:post_x_i}), or
its numerical version (eq.~\ref{eq:post_indiv_uniform_grid}). A more
efficient sweep of the parameter space would be to use a sampling
scheme which traces the full probability distribution
(eq.~\ref{eq:full_posterior}). One such `importance sampling' scheme
is the Metropolis algorithm (Metropolis \etal 1953), which samples
preferentially regions where the posterior is large. This is the
parameter-space exploration method adopted in this work.

Our implementation of the Metropolis sampler was as follows. Starting
from an initial arbitrary point, at each iteration $s$ we pick one of
the 13 variables at random and change it by an uniform deviate ranging
from $-\epsilon$ to $+\epsilon$, producing a new state $s+1$. If the
variable is one of the 12 $x_i$'s, the whole set is renormalized to
unit sum. Moves towards unphysical values ($x_i < 0$ or $> 1$, $A_V <
0$) were truncated. Downhill moves (i.e., towards smaller $\chi^2$)
are always accepted, whilst changes to less likely states are accepted
with probability $e^{- ({\cal E}_{s+1} - {\cal E}_s)}$, thus avoiding
trapping onto local minima. This scheme, widely employed in
statistical mechanics (Press \etal 1992; MacKay 2001 and references
therein), produces a distribution which tends to the correct one as
the number of samples $N_s$ increase. There is, however, no universal
prescription to choose optimal values for $\epsilon$ or $N_s$, so some
experimentation is needed.

\subsection{Performance tests}

\label{sec:Performance_tests}

In order to test our Metropolis EPS code we performed a series of
simulations for artificial galaxies generated out of the base. Two
sets of test galaxies were used. The first is composed of the 26
galaxies used by S91 (see their Table 5) to test their minimization
algorithm. As they did not synthesize colors, we have used a fixed
value of $A_V = 2/3$ to redden all their galaxies. The second set is
composed of 100 test-galaxies whose population vectors and extinctions
were generated randomly using a scheme which insures that in many
cases a few components dominate the light. All simulations presented
in this paper were performed sampling $N_s = 10^8$ states, with the
`visitation parameter' $\epsilon$ set to 0.005 for both the $x_i$'s
and $A_V$. Experiments were also performed changing these quantities,
and found to give nearly identical results.  We note that less samples
can be used if one is only interested in the mean (${\bf x},A_V$), as
this naturally converges much faster than its full probability
distribution. Simulations were performed for severals values of
$(S/N)_{5870}$ and different sets of observables. Here we restrain the
discussion to the illustration of the sampling method.

%***FIG***FIG***FIG***FIG***FIG***FIG***FIG***FIG***FIG***FIG***FIG
\begin{figure*}
%\resizebox{\textwidth}{!}{\includegraphics{BWFig_illustrate_method.eps}}
\resizebox{\textwidth}{!}{\includegraphics{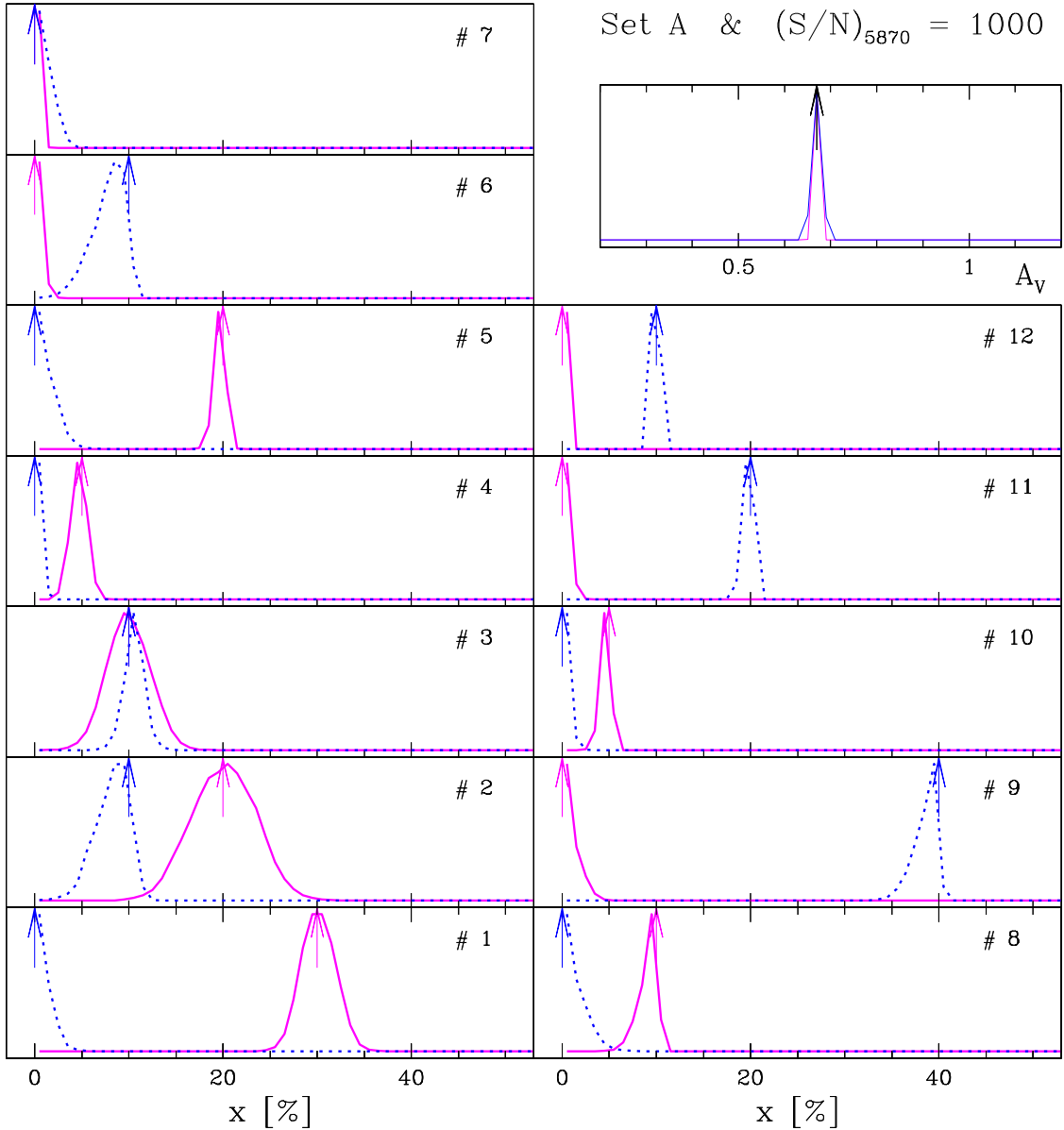}}
\caption{Individual posterior probability distributions for each of
the components of the population vector ${\bf x}$ and the extinction,
computed for an old population dominated test galaxy (solid lines) and
a `younger' galaxy (dotted lines). The top arrows indicate the input
values of the corresponding parameter in each panel. The numbering of
the ${\bf x}$ components follows the convention in Table 1. All 16
observables in Bica's base (9 absorption lines $+$ 7 continuum points)
were synthesized in this example and a S/N of 1000 was adopted to
illustrate the convergence of the Metropolis sampler in the limit of
(almost) perfect data.}
\label{fig:Test_Metropolis}
\end{figure*}
%***FIG***FIG***FIG***FIG***FIG***FIG***FIG***FIG***FIG***FIG***FIG

In Fig.~\ref{fig:Test_Metropolis} we plot the individual projected
posterior probability distributions for each of the 12 $x_i$'s plus $A_V$
for two of S91's test galaxies composed of widely different stellar
populations.  The posteriors were computed synthesizing all 9 lines and 7
colors in the base (`Set A', as defined below) and using $(S/N)_{5870} =
1000$ to define the errors in the observables. This is clearly an
unrealistic, highly idealized situation, but it serves as a test-bed for
the method in the limit of perfect data. The simulations were started from
the $x_i = 1/12$ and $A_V = 0.5$ point. The plot shows that the posterior
probabilities are sharp functions peaking at the expected input value,
marked by upper arrows, demonstrating that the Metropolis sampler converges
to the most likely region.  One notes, however, that the strong components
in the old age bin (such as $x_1$, $x_2$ and $x_3$ in the `old galaxy'
example, drawn as solid lines) have broader individual posteriors. This
effect is further discussed in the next section.

In the opposite limit of large errors, $\sigma \rightarrow \infty$
(the `infinite temperature' regime, as it would be called in
statistical mechanics) the likelihood loses its discriminating power
and becomes $\sim$ constant over the whole parameter-space.
Eq.~(\ref{eq:post_x_i}) then tends to $P(x_i) \propto (1 - x_i)^{10}$
for a $n_\star = 12$ base.  In this regime the data does not constrain
the model at all, and this expression for $P(x_i)$ simply reflects the
volume of the ${\bf x}$-space spanned by a given value of $x_i$. Not
surprisingly, this distribution peaks at $x_i = 0$, and has a mean
value $\ov{x_i} = 1/12$. We have verified that the Metropolis scheme
recovers this distribution satisfactorily as $S/N \rightarrow 0$,
which serves as a further test of the adequacy of this sampling
method. The skewing of several of the individual posteriors plotted in
Figs.~\ref{fig:Errors_in_observables} and
\ref{fig:Sets_of_observables} towards small $x_i$ is partly due to
this effect, which drags the components to an intrinsically larger
region of ${\bf x}$-space.

%***FIG***FIG***FIG***FIG***FIG***FIG***FIG***FIG***FIG***FIG***FIG
\begin{figure*}
%\resizebox{\textwidth}{!}{\includegraphics{BWFig_MET_steps.eps}}
\resizebox{\textwidth}{!}{\includegraphics{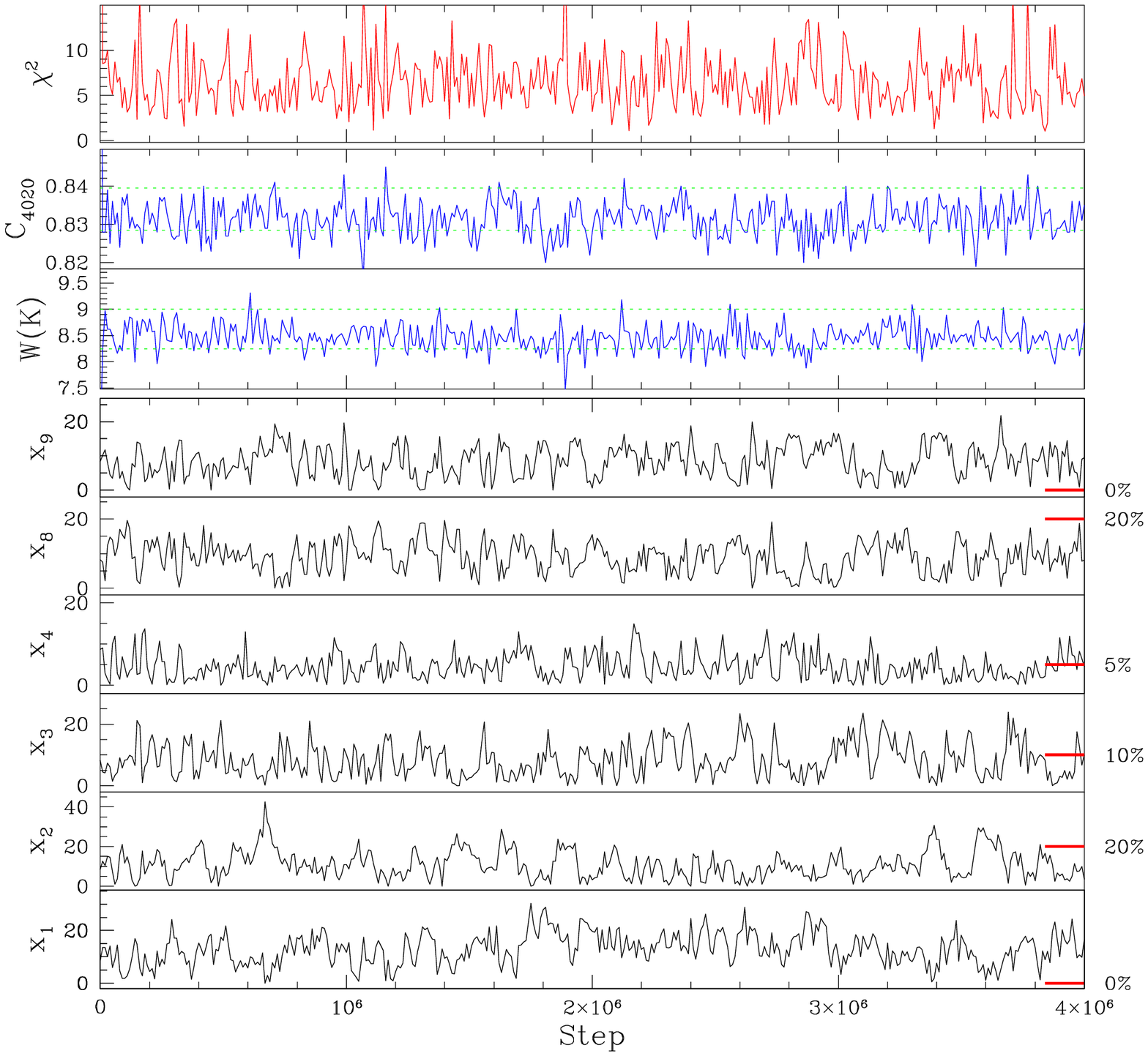}}
\caption{Illustration of the `step by step' Metropolis walk through
the parameter space. The six bottom panels illustrate the evolution of
selected components of the population vector (in \%), whose input
$x_i$ values are indicated at the right edge of each plot. The values
of the equivalent width of CaIIK and of the color $C_{4020}$ are
plotted for each step in the third and second panels from the top. The
dotted horizontal lines in these plots mark the $\pm 1 \sigma$ range
around the observed value. The top panel shows the $\chi^2$ for each
sampled state. For clarity, only the first few million steps are show,
and only 1 for every 10000 sampled states is actually plotted. All 16
observables in the base and $S/N = 60$ were used in this example.}
\label{fig:Metropolis_steps} 
\end{figure*}
%***FIG***FIG***FIG***FIG***FIG***FIG***FIG***FIG***FIG***FIG***FIG

The evolution of the Metropolis walk through the 13-D (${\bf x},A_V$)
space is illustrated in Fig.~\ref{fig:Metropolis_steps}. For clarity,
only 6 components of ${\bf x}$ and just one state for each $10^4$
steps are shown. Note that projected $P(x_i)$ posteriors such as those
in Fig.~\ref{fig:Test_Metropolis} are essentially histograms of the
$x_i$ states sampled in walks as that in
Fig.~\ref{fig:Metropolis_steps}, but weighted by their $\exp
-\chi({\bf x})^2/2$ likelihood. The input values of the components
shown are indicated at the right edge of the plot for comparison. This
particular example corresponds to a simulation with $S/N = 60$ and
with all 16 observables synthesized.  One sees relatively large
excursions of the parameters from their true values. Furthermore,
anti-correlations between some of the components ($x_1$ and $x_2$,
$x_8$ and $x_9$) are clearly visible. This `mirror effect' becomes
much more pronounced as $S/N$ increases, whereas for low $S/N$ the
broadening of the likelihood function washes out such correlations.
This is a consequence of redundancies between several of the base
components, as further discussed below. The top panels in
Fig.~\ref{fig:Metropolis_steps} show the evolution of the $\chi^2$ and
how two of the synthesized observables, $W$(CaIIK) and $C_{4020}$,
oscillate very much within their respective $\pm 1$ sigma error
ranges. One therefore anticipates an excellent fit to the data, and
indeed the $\chi^2$ obtained for the {\it mean} (${\bf x},A_V)$
`solution' in this example is just 0.8 (see also
Fig.~\ref{fig:input_X_output_observables}).

Overall, we conclude that the Metropolis algorithm is an efficient
sampler of the parameter space for EPS problems. It provides a much
more continuous mapping of ${\bf x}$ and $A_V$ than would be possible
with a uniform grid. Furthermore, it is substantially more efficient
computationally. A uniform grid with $10^8$ points would only sample
each $x_i$ at $\sim 7\%$ steps and 10 values for $A_V$. By
concentrating on the relevant regions of the parameter space, we
achieve a much better resolution, as can be judged by the smoothness
of the posterior distributions on scales of 1--2\%.

One possible variation of this scheme is to combine the uniform grid
and Metropolis routines, starting a Markov chain from each point on a
uniform grid. Experiments with this alternative scheme were performed,
and found to produce essentially the same results. This is to be
expected, since we verified that the starting point has little effect
upon the resulting posterior distributions. In fact, we performed a
series of simulations using the {\it input parameters as the starting
point} and obtained identical results.  Though further improvements
are possible, the simple Metropolis sampler already provides a
substantial improvement over previous works with this base.

\section{Tests of the base}

\label{sec:Test_of_Bicas_Base}

Having developed a probabilistic formulation of EPS and a new sampling
method, we now present a series of numerical experiments designed to
evaluate the actual ability to recover the stellar population mix of
galaxies using Bica's base. The tests were performed with the same
series of fictional galaxies described in section
\ref{sec:Performance_tests}. Simulations were performed for
$(S/N)_{5870} = 10$, 30, 60, 100, 300 and 1000, and three different
sets of observables:

\begin{itemize}

\item Set A: All 16 observables in the base.

\item Set B: Only the 9 absorption line equivalent widths.

\item Set C: The $W$'s of CaII~K, CN, G-band and MgI+MgH plus the
continuum fluxes at 3660, 4020, 4510 and 6630, all normalized to 5870
\AA.

\end{itemize}

Set A is the ideal, as it uses all information in the base. Set B is
used as a test of how much is actually gained by synthesizing colors as
well as $W$'s, while Set C is composed of the observables used by Cid
Fernandes \etal (1998) and Schmitt \etal (1999) to characterize the
stellar content of active galaxies and their hosts through long-slit
spectroscopy. These data will be used to perform a spatially resolved
EPS study in a future communication.

\subsection{Effects of the errors in the observables}

\begin{figure*}
%\resizebox{\textwidth}{!}{\includegraphics{BWFig_effects_of_ERRORS.eps}}
\resizebox{\textwidth}{!}{\includegraphics{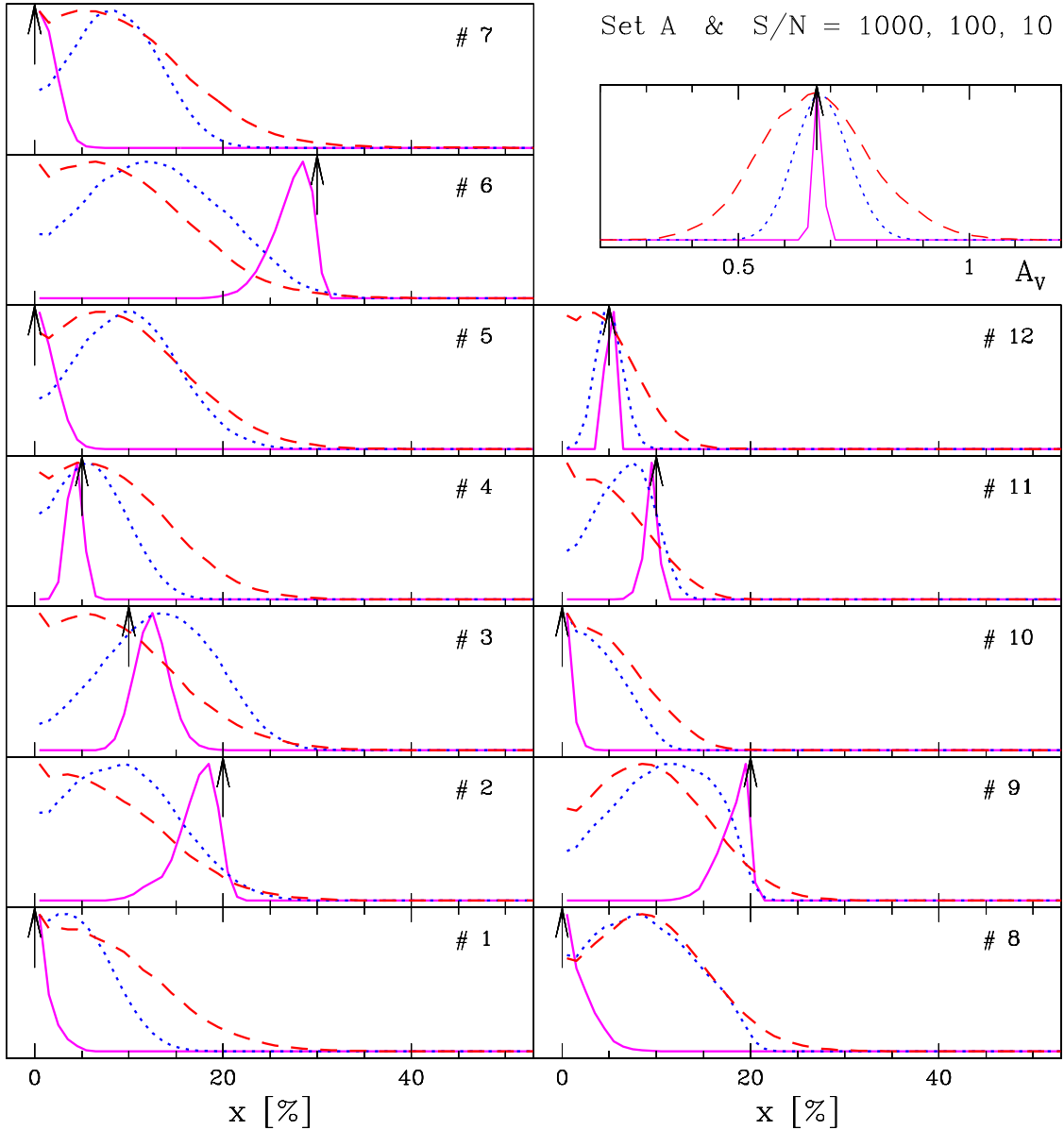}}
\caption{Effect of $S/N$ upon the individual probability distributions
of the synthesis parameters for Set A observables. The different lines
correspond to $S/N = 1000$ (solid), 100 (dotted) and 10
(dashed). Notice how the probability of ${\bf x}$ components with a
large contribution (like $x_2$, $x_6$ and $x_9$ in this example) are
progressively redistributed among other components as the noise
increases.}
\label{fig:Errors_in_observables}
\end{figure*}

We first investigate the effects of the measurement uncertainties in the
observables upon the results of the synthesis. Qualitatively, one expects
the degradation of the data quality to {\it broaden} the probability
distributions. Furthermore, a {\it shift} and {\it skewing} of the
$P(x_i)$'s of intrinsically large components towards small $x_i$'s is also
expected for low $S/N$ due to intrinsically larger number of states in this
region of ${\bf x}$ (the `dragging effect' discussed in
\S\ref{sec:Performance_tests}). This is exactly what is observed in
Fig.~\ref{fig:Errors_in_observables}, where we overplot the individual
posteriors for $S/N = 1000$, 100 and 10 for one particular test galaxy of
S91, synthesized with Set A observables.  As in
Fig.~\ref{fig:Test_Metropolis}, the true parameters are indicated by the
top arrows. The broadening of posteriors in this case can be safely
attributed to the errors alone, since, as demonstrated by Pelat (1998) most
of the test galaxies in S91 (including the one in
Fig.~\ref{fig:Errors_in_observables}) have a unique algebraic solution in a
$W$-only EPS problem, despite the fact that the number of degrees of
freedom (11 if we do not model the colors) falls short of the number of
observables (9 $W$'s).

%***FIG***FIG***FIG***FIG***FIG***FIG***FIG***FIG***FIG***FIG***FIG
%***FIG***FIG***FIG***FIG***FIG***FIG***FIG***FIG***FIG***FIG***FIG

%***FIG***FIG***FIG***FIG***FIG***FIG***FIG***FIG***FIG***FIG***FIG
\begin{figure*}
%\resizebox{\textwidth}{!}{\includegraphics{BWFig_CovMat.eps}}
\resizebox{\textwidth}{!}{\includegraphics{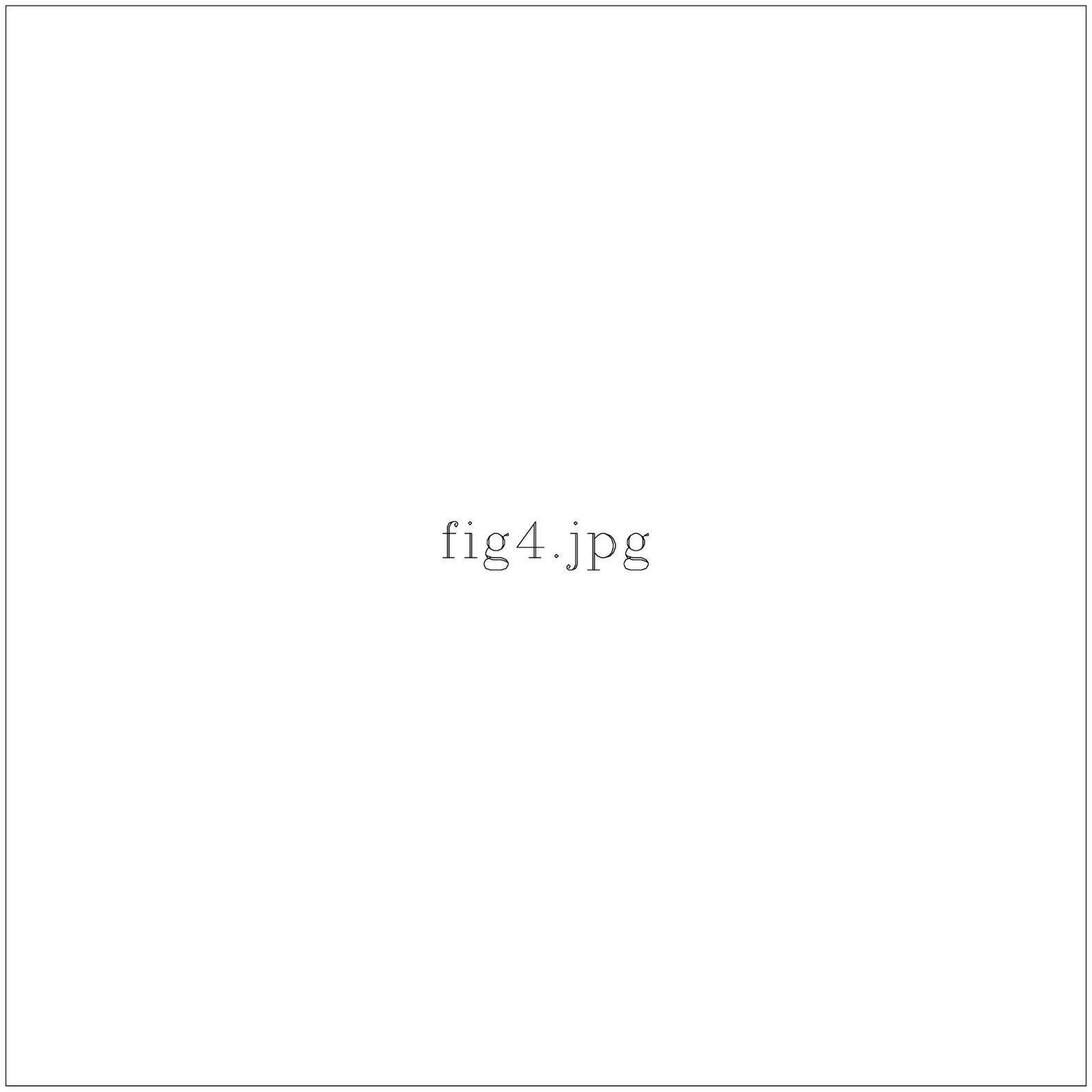}}
\caption{`Correlation Matrix' for the test galaxy in
Fig.~\ref{fig:Errors_in_observables}. Each $x_i \times x_j$ panel
shows one every 10000 of the first $10^7$ steps of a Metropolis run
for $(S/N)_{5870} = 300$ and Set A observables. The values at the
bottom of each row indicate the input value of the corresponding
component ($x_1 = 0$, $x_2 = 20\%$, \ldots $x_{11} = 10\%$, $x_{12} =
10\%$). These are also indicated by the solid circles in each
panel. Small and big tick marks are spaced by 2 and 10\% respectively
in all plots. Note the strong anti-correlations between adjacent
components of same age.}
\label{fig:Correlation_Matrix}
\end{figure*}
%***FIG***FIG***FIG***FIG***FIG***FIG***FIG***FIG***FIG***FIG***FIG

Despite the large number of constraints in Set A, one sees that the
individual posterior distributions are substantially broad even for a $S/N
= 100$ spectrum.  The average standard deviation of the $x_i$'s for this
combination of $S/N$ and observables is 3\% over all simulations, but can
reach more than 10\% in individual components.  Furthermore, the dominant
components, 2, 6 and 9 in this example, are all badly affected by the
errors. Fig.~\ref{fig:Errors_in_observables} shows that decreasing data
quality progressively shifts $P(x_2)$ towards smaller values, whereas the
opposite happens with $x_1$. This {\it compensation} is also seen between
components in the 1 Gyr ($x_5$, $x_6$ and $x_7$), 100 Myr ($x_8$ and
$x_9$), and 10 Myr ($x_{10}$ and $x_{11}$) age bins and is the same
`mirror' effect observed before in Fig.~\ref{fig:Metropolis_steps}. A
compelling visualization of these compensations is given in
Fig.~\ref{fig:Correlation_Matrix}, where we plot the individual Metropolis
states for the test galaxy employed in Fig.~\ref{fig:Errors_in_observables}
in a $x_i \times x_j$ matrix.  The $S/N$ of 300 in this plot was chosen for
clarity purposes; the same structure is present for other values, with the
scatter around the strongest correlations increasing steadily as $S/N$
decreases, to the point that at $S/N = 10$ the anti-correlations between
components 1 and 2, or 8 and 9 become clouds similar to $x_1
\times x_5$ in Fig.~\ref{fig:Correlation_Matrix}. All anticorrelations
identified in Fig.~\ref{fig:Errors_in_observables} ($x_1 \times x_2$,
$x_6 \times x_5$, etc.) are clearly represented in this graphical
version of the correlation matrix.

\begin{figure*}
%\resizebox{\textwidth}{!}{\includegraphics{Fig_SynBase.eps}}
\resizebox{\textwidth}{!}{\includegraphics{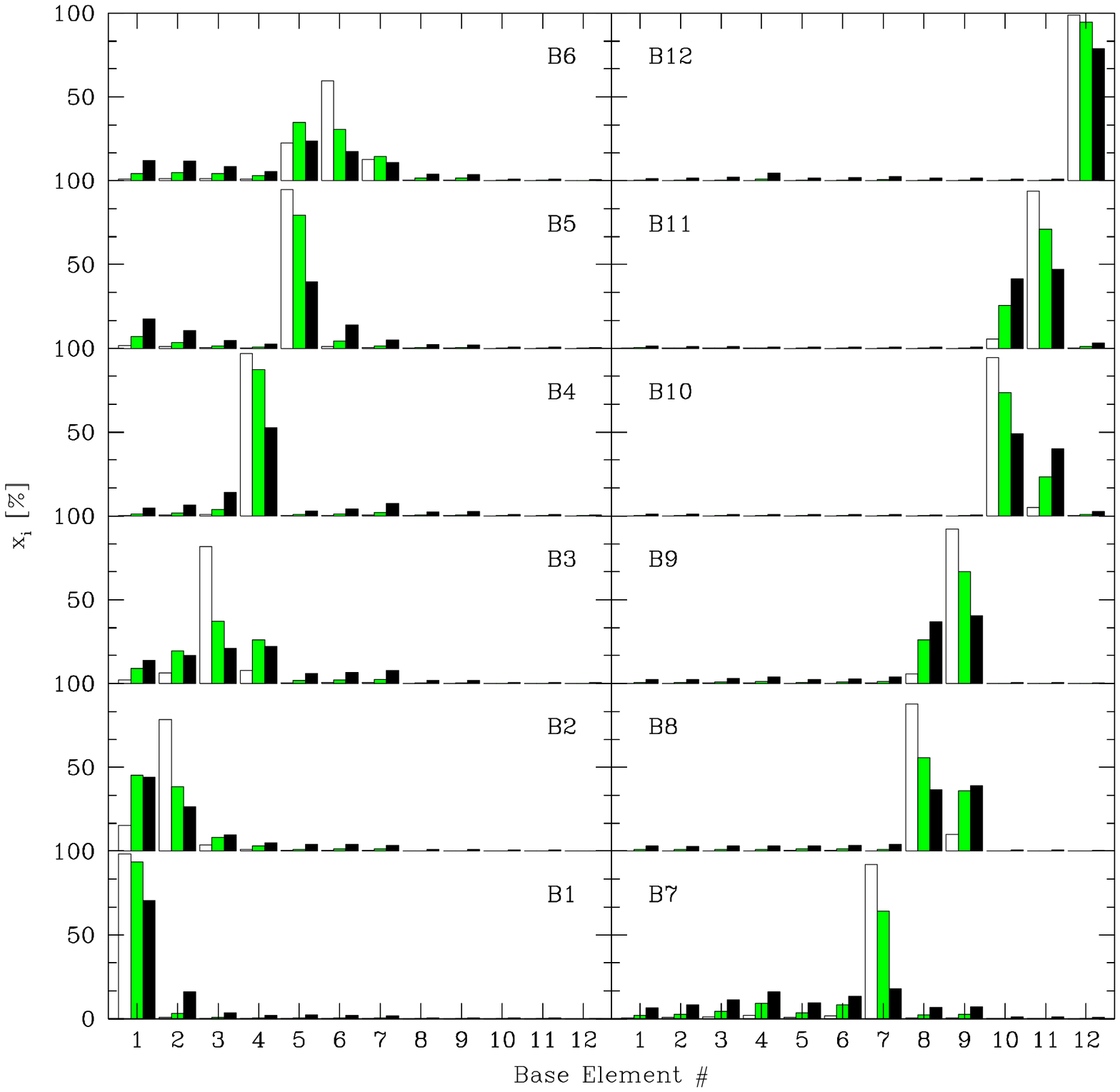}}
\caption{Synthesis of the base elements. Each panel shows the mean
synthetic population vector (vertical axis) for 12 test galaxies whose 16
observables were generated with $x_i = 100\%$, $i = 1,\ldots 12$ (labeled
B1\ldots B12 respectively). Empty, shaded and filled histograms correspond
to $S/N = 100$, 30 and 10 respectively. Ideally, all histograms should be
concentrated in the component used to generate the observables. In
practice, some bases elements, most notably 2, 3 and 6, are well
synthesized by combinations of others even for large $S/N$.}
\label{fig:SyntBase}
\end{figure*}

These effects are rooted in the internal structure of the base.  By
definition, the base elements must be linearly independent, and the S91
base complies with this algebraic condition. Yet, some of its elements are
{\it practically} linearly dependent, in the sense that combinations of
other elements can recover them to a high degree of accuracy.  As the noise
increases, the residuals between representing an element by its exact $W$'s
and $C$'s and a combination including other elements with more extreme
values for the observables become statistically insignificant, explaining
the compensations detected above.  To demonstrate this we synthesized the
base elements themselves with our code. Results for Set A observables and
three different values of $S/N$ are shown in Fig.~\ref{fig:SyntBase}. Each
panel corresponds to one of these 12 extreme test galaxies (labeled
B1\ldots B12), with the {\it mean} synthetic values of each of the
components plotted along the vertical axis. The empty, shaded and filled
histograms correspond to $S/N = 100$, 30 and 10 respectively. Some of the
components, most notably numbers 2, 3 and particularly 6, are synthesized
with large contributions of others even for $S/N = 100$. For the $x_6 =
100\%$ model (top left panel of Fig.~\ref{fig:SyntBase}), for instance, we
find $\ov{x_6} = 60\%$ for this high $S/N$, with 35 of the remaining 40\%
redistributed among components 5 and 7. Yet, the 16 observables are very
well reproduced by this combination, with a total $\chi^2$ of just
3.8. This spreading of the light fractions becomes more pronounced for
lower $S/N$'s, to the point that at $S/N = 10$, components like 8 and 9, or
10 and 11 cannot be distinguished anymore, and end up dividing their
contributions in roughly equal shares, with residuals spread over other
components.  The spread is much smaller in the synthesis of components 1,
4, 5 and 7, which represent extreme metallicities within age groups. For
intermediate $Z$ systems (models B2, B3 and B6), however, these extreme
components attract much of the percentage light contribution spilled over
from neighboring populations of same age, so that their individual
contributions in a real stellar population mix are as unreliable as the
others.

This experiment demonstrates that for practical purposes the base is still
linearly dependent, at least in a statistical sense, despite the effort of
S91 in reducing the highly redundant original base of B88.  This
`noise-induced', or `statistical linear dependence', as we may call it,
could indeed be inferred from the PCA analysis of S91, which showed that
very little of the variance is associated with the last few
eigenvectors. Within our probabilistic formulation, this `statistical
linear dependence' defines the structure of the likelihood function, which
spreads first along directions in ${\bf x}$-space producing similar
observables, carrying the mean ${\bf x}$ to more densely populated
regions. This effect explains the redistribution of the probability
observed in Figs.~\ref{fig:Errors_in_observables},
\ref{fig:Correlation_Matrix} and all other simulations.

%***FIG***FIG***FIG***FIG***FIG***FIG***FIG***FIG***FIG***FIG***FIG
%***FIG***FIG***FIG***FIG***FIG***FIG***FIG***FIG***FIG***FIG***FIG

%***FIG***FIG***FIG***FIG***FIG***FIG***FIG***FIG***FIG***FIG***FIG
\begin{figure*}
%\resizebox{\textwidth}{!}{\includegraphics{BWFig_inXout_individual_ave_E.eps}}
\resizebox{\textwidth}{!}{\includegraphics{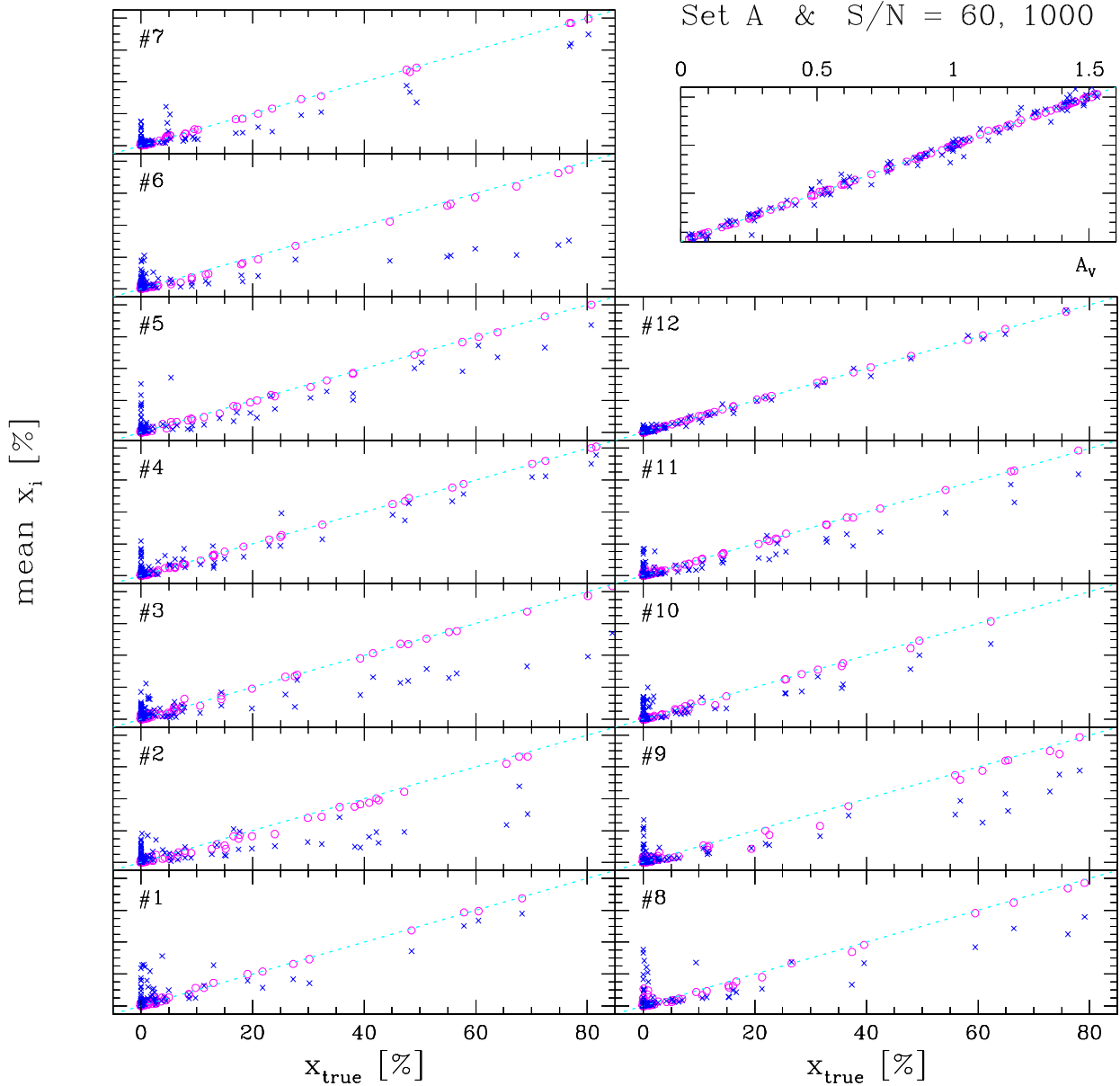}}
\caption{Input $\times$ output plots for all 13 individual parameters
in the synthesis. Open circles map the results for Set A and $S/N =
1000$, while crosses correspond to Set A and $S/N = 60$. The output
parameters are the {\it mean} values sampled from the global
likelihood function. The dotted, diagonal lines in each panel mark the
`input = output' line. Error bars on the model parameters are not
shown for clarity. Notice the systematic underestimation of large
$x_i$'s and the corresponding overestimation of weak proportions for
$S/N = 60$.  }
\label{fig:individual_input_X_ouput}
\end{figure*}
%***FIG***FIG***FIG***FIG***FIG***FIG***FIG***FIG***FIG***FIG***FIG

%***FIG***FIG***FIG***FIG***FIG***FIG***FIG***FIG***FIG***FIG***FIG
\begin{figure*}
%\resizebox{\textwidth}{!}{\includegraphics{BWFigPaper_inXout_OBSERVABLES.eps}}
\resizebox{\textwidth}{!}{\includegraphics{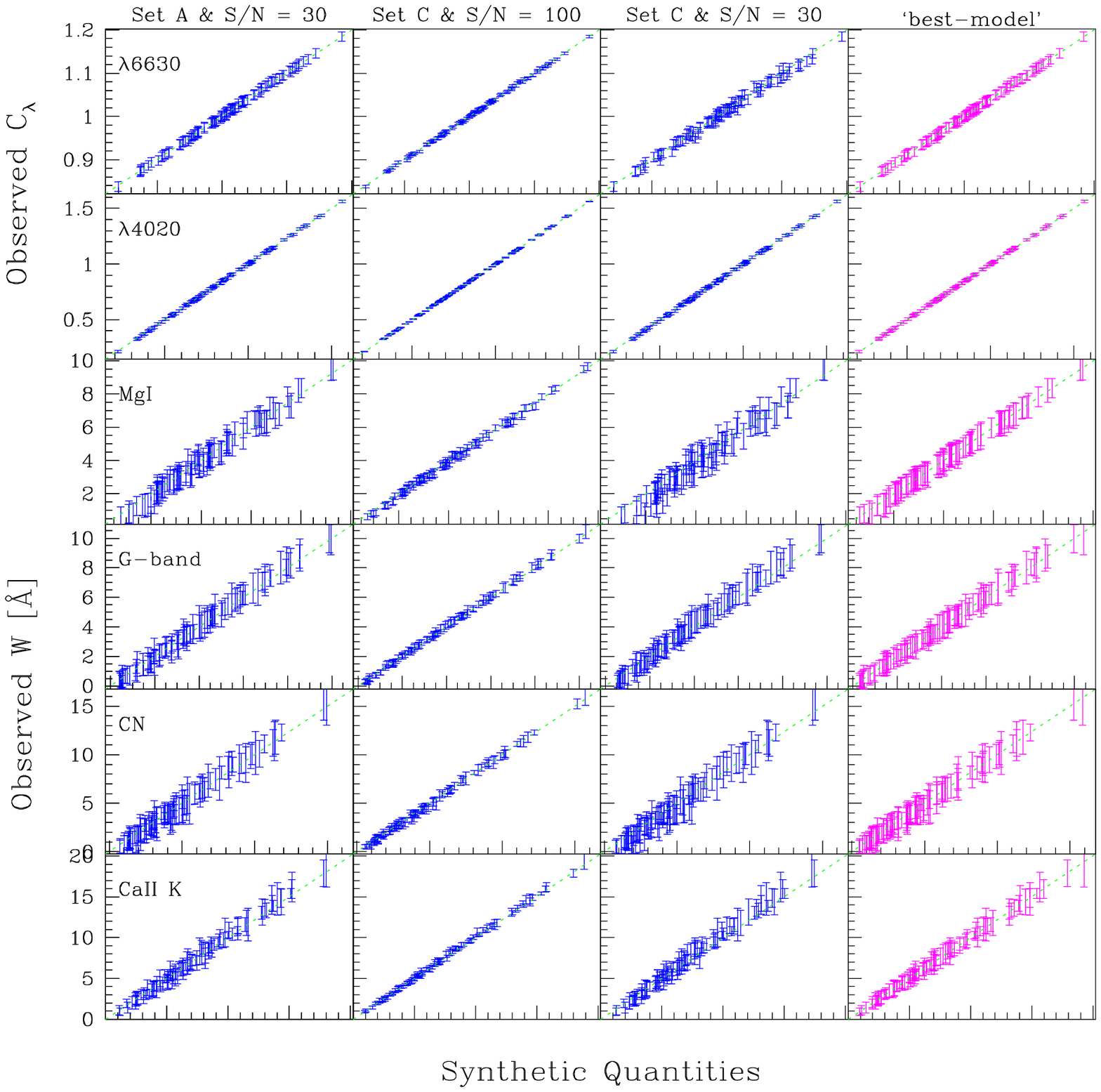}}
\caption{Input $\times$ output observables for a series of 100
test-galaxies synthesized with the Metropolis-EPS sampler. The three
left columns of plots correspond to the three indicated combinations
of the set of observables and $S/N$. For these plots the x-axis
quantities were computed with the {\it mean} (${\bf x},A_V$)
solution. The plots on the right column were made with Set C and $S/N
= 30$ (as those in the third column), but the synthetic observables in
this cases were computed with the best model found during the
Metropolis runs. Only the equivalent widths of four absorption
features and two colors exhibited, but results for the remaining
observables are equally good. The vertical and horizontal scales are
the same for each panel, and the input $=$ output is indicated by a
dotted line. Notice the excellent agreement, despite the fact that the
input {\it parameters} are not accurately recovered by the synthesis
process.}
\label{fig:input_X_output_observables}
\end{figure*}
%***FIG***FIG***FIG***FIG***FIG***FIG***FIG***FIG***FIG***FIG***FIG

It is therefore extremely difficult to retrieve accurately all the input
stellar population parameters even for excellent data.  This is further
illustrated in Fig.~\ref{fig:individual_input_X_ouput}, where the input
parameters are plotted against their mean values, as derived from the
Metropolis runs. One hundred artificial galaxies are plotted in each panel,
with open circles corresponding to Set A observables and $S/N = 1000$, and
crosses indicating the results for Set A and $S/N = 60$.  A systematic
underestimation of strong components is seen for $S/N = 60$, a good
spectrum by any standard. This happens at the expense of an overestimation
of the weaker components, which produces the large scatter seen in the
bottom left corners of Fig.~\ref{fig:individual_input_X_ouput}. The average
uncertainties in the $x_i$'s are of order $\sigma(x_i) \sim 3$--4\% for
$S/N = 60$. These are not enough to account for the large differences
between input and output $x_i$'s depicted in
Fig.~\ref{fig:individual_input_X_ouput}, so the errors induce a true bias
in $\ov{\bf x}$. Note also that, in accordance with the discussion above,
this bias is smaller for models with strong contributions of components 1,
4, 5 and 12, because of their extreme locations in the space of
observables.

Only for unrealistically large $S/N$'s, which far exceed the quality
of the data used to build the base in the first place, one is able to
break the `statistical linear dependence' of the base by
distinguishing fine details in the observables. This should {\it not}
be interpreted as a failure of the method, as what the code actually
synthesizes are the observables!  These are {\it very precisely}
reproduced by the {\it mean} (${\bf x}, A_V$), as illustrated in
Fig.~\ref{fig:input_X_output_observables} for the $W$'s of CaII K, CN,
G-band and MgI and two continuum colors. The remaining observables,
not show in the figure, are equally well reproduced. Furthermore,
whilst $S/N > 300$ is needed to recover accurately the model
parameters from the mean solution, the agreement between the synthetic
and measured observables is excellent for {\it any} $S/N$.  Obviously,
an even better agreement is obtained if instead of $(\ov{\bf
x},\ov{A_V})$ the best model sampled during the Metropolis excursion
is used to reconstruct the observables, as illustrated in the fourth
column of plots in Fig.~\ref{fig:input_X_output_observables} (see
\S\ref{sec:Best_X_Mean}).

In conclusion, this comparison of input and output observables shows
that the spreading of the probability and the confusion between
components seen in Figs.~\ref{fig:Errors_in_observables} and
\ref{fig:SyntBase} is not an artifact of the method, but a consequence
of the internal structure of the spectral base.  As found in other EPS
studies (O'Connell 1996 and references therein) synthesizing the
observables is one thing, but trusting the detailed star-formation
history and chemical evolution implied by the synthesis is an
altogether different story. On the positive side, a notable fact about
Fig.~\ref{fig:SyntBase} is that, consulting Table~\ref{tab:Base}, one
realizes that the reshuffling of the strength of the components occurs
preferentially among populations of same {\it age}, an effect also
clearly seen in Fig.~\ref{fig:Correlation_Matrix}. Though some
redistribution among base elements of different age and $Z$ in the
older 1 and 10 Gyr bins due to the age-$Z$ degeneracy also occurs,
this indicates that the age distribution may be well recovered by the
synthesis (\S\ref{sec:Grouped_Results}).

\subsection{Effects of different sets of observables}

%***FIG***FIG***FIG***FIG***FIG***FIG***FIG***FIG***FIG***FIG***FIG
\begin{figure*}
%\resizebox{\textwidth}{!}{\includegraphics{BWFig_effects_of_OBSERVABLES.eps}}
\resizebox{\textwidth}{!}{\includegraphics{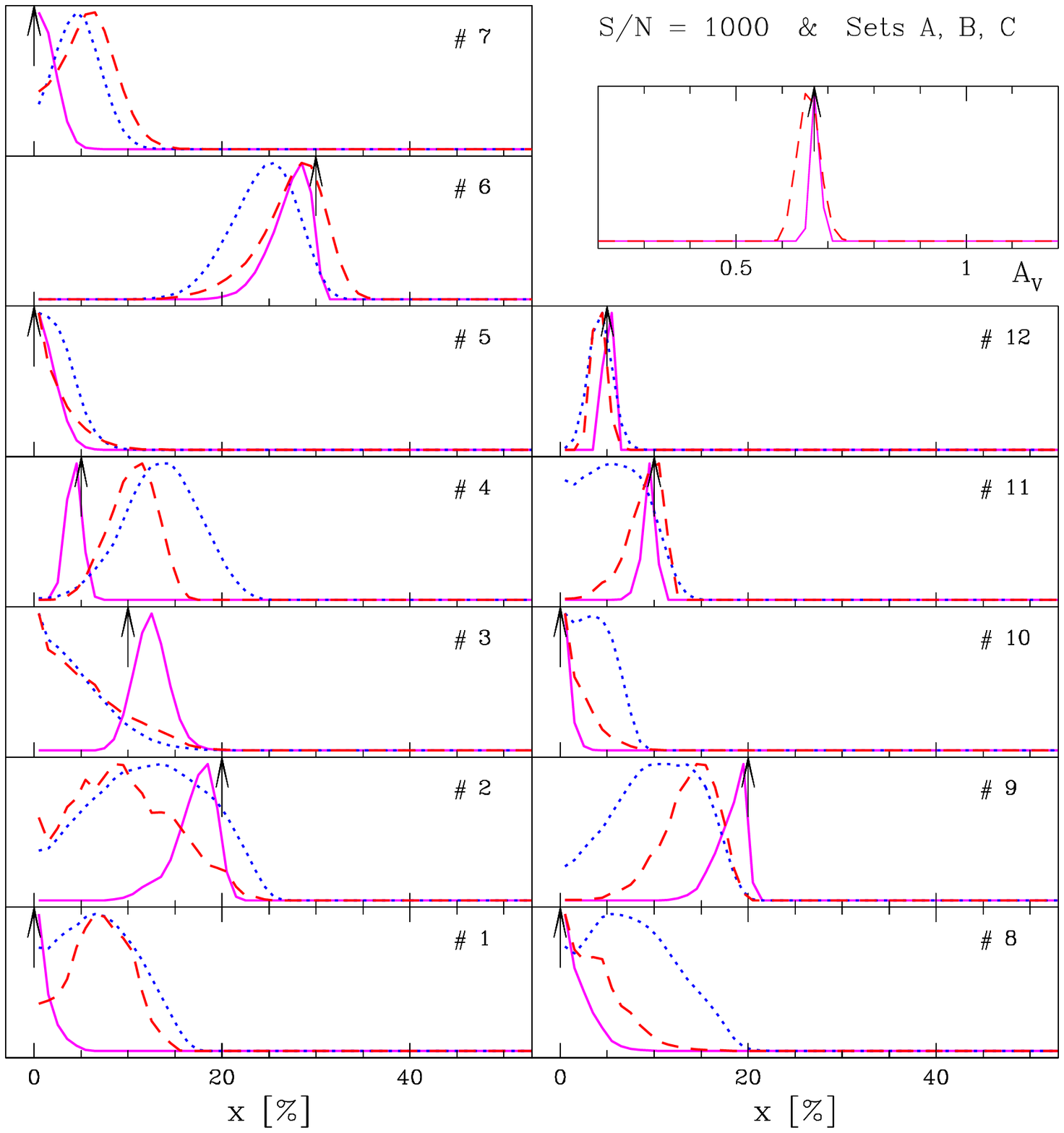}}
\caption{Analogous to Fig.~\ref{fig:Errors_in_observables}, but
exploring the effect of different sets of observables upon the
synthesis parameters.  Solid lines correspond to Set A (all
observables), dotted lines to Set B (only 9 equivalent widths) and
dashed lines to Set C (4 lines and 4 colors).}
\label{fig:Sets_of_observables}
\end{figure*}
%***FIG***FIG***FIG***FIG***FIG***FIG***FIG***FIG***FIG***FIG***FIG

The effects of which set of observables is used in the synthesis are
illustrated in Fig.~\ref{fig:Sets_of_observables}, where results for
Sets A, B and C for the same test galaxy as in
Fig.~\ref{fig:Errors_in_observables} and $S/N = 1000$ are
compared. The figure shows that it is very advantageous to synthesize
colors along with equivalent widths, despite the fact that one
increases the dimensionality of the problem with the inclusion of
$A_V$. This can be seen by comparing the performances of Sets A and
B. The information contained in the colors not only improves the
estimation of ${\bf x}$ but also allows a very good determination of
$A_V$, which, unlike the population vector, is {\it always} well
recovered by the synthesis.

Fig.~\ref{fig:Sets_of_observables} also shows how important it is to
provide as many observational constraints as possible to constrain the
synthesis, as Set A recovers the parameters much better than Sets B and
C. Decreasing the number of observables in the synthesis thus have the same
overall effect as decreasing the $S/N$. The width of the posteriors for Set
C are partially attributed to the algebraic degeneracy in this set (8
observables and 12 degrees of freedom), which implies the existence of
exact solutions in a sub-space of (${\bf x}$,$A_V$). That is not the case
of Set A for the test galaxy in this example (Pelat 1998), but there, as
for other sets, the `statistical linear dependence' of the base is very
efficient in spreading the likelihood even for such an idealized S/N
ratio. The excellent agreement between the `observed' and synthesized
observables even for much worse S/N
(Fig.~\ref{fig:input_X_output_observables}) illustrates this point.
Algebraic degeneracy is not as critical as the `statistical degeneracy'
induced by the combined effects of noise, limited data and the internal
correlations in the base.

Studies which synthesize only $W$'s sometimes use the resulting
population vector to compute a predicted continuum shape and derive
$A_V$ {\it a posteriori} through the comparison with the observed
continuum (e.g., B88; Boisson \etal 2000). From our Set B simulations,
we find that this procedure yields values of $A_V$ 2--3 times less
accurate than obtained with Set A. For $S/N = 60$, for instance, Set A
yields an rms dispersion in $A_V$ of 0.06 mag around the true value,
whereas the dispersion for set B is 0.13 mag. Since this is a
relatively small difference, these experiments validate the {\it a
posteriori} computation of the extinction. Colors are more useful to
constrain the population vector than to estimate $A_V$.

\subsection{Age grouped results}

\label{sec:Grouped_Results}

%***FIG***FIG***FIG***FIG***FIG***FIG***FIG***FIG***FIG***FIG***FIG
\begin{figure*}
%\resizebox{\textwidth}{!}{\includegraphics{BWFig_inXout_AGEgrouped_ave_E.eps}}
\resizebox{\textwidth}{!}{\includegraphics{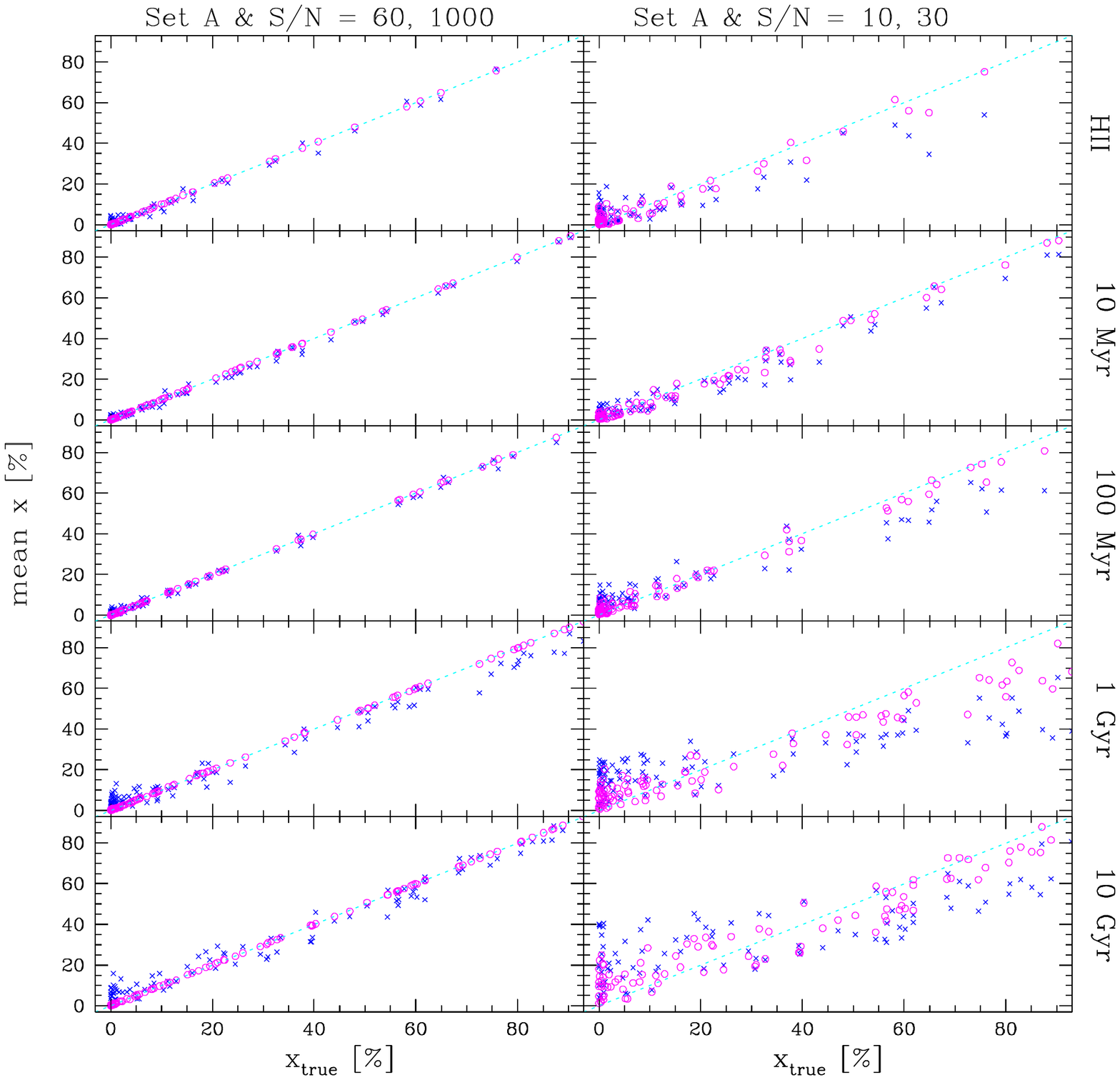}}
\caption{ Age-binned light fractions in 100 input test galaxies versus
the corresponding synthetic means.  {\it Left panels:} Circles
correspond to $S/N = 60$, and crosses to $S/N = 1000$, as in
Fig.~\ref{fig:individual_input_X_ouput}. {\it Right panels:} Circles
correspond to $S/N = 30$, and crosses to $S/N = 10$. All results are for
Set A observables.
}
\label{fig:Age_and_Z_grouped_input_X_ouput}
\end{figure*}
%***FIG***FIG***FIG***FIG***FIG***FIG***FIG***FIG***FIG***FIG***FIG

The results of the tests above reveal a striking difficulty to
accurately determine all 12 base components of Bica's base in the
presence of even very modest noise and/or when not all base
observables are available for the synthesis. As indicated by the
simulations, both the measurement errors and the use of reduced sets
of observables act primarily in the sense of spreading a strong
contribution in one component preferentially among base elements of
same age. {\it Grouping} the population vector in age bins should thus
produce more robust results.  This expectation is confirmed in
Fig.~\ref{fig:Age_and_Z_grouped_input_X_ouput}, where, analogous to
what was done in Fig.~\ref{fig:individual_input_X_ouput}, we plot the
input values of the $x_i$'s against the output $\ov{x_i}$'s, but now
for the five age binned groups, also for Set A simulations.  The left
panels correspond to the same values of $S/N$ used in
Fig.~\ref{fig:individual_input_X_ouput}, so that one can appreciate
the enormous improvement achieved by grouping equal age elements.  The
right panels show that good results are also obtained for $S/N$ of 30
(circles), and that young populations ($\le 100$ Myr) are reasonably
well traced even for $S/N$ as low as 10 (crosses).  Though some
deviations survive, specially for the smaller $S/N$'s and the older
age bins, the age-binned synthesis results are much more reliable than
the results for the individual components, and, most importantly, do
{\it not} require unrealistic $S/N$.

\begin{figure*}
%\resizebox{\textwidth}{!}{\includegraphics{BWFig_inXout_ave_AGE_A_and_D.eps}}
\resizebox{\textwidth}{!}{\includegraphics{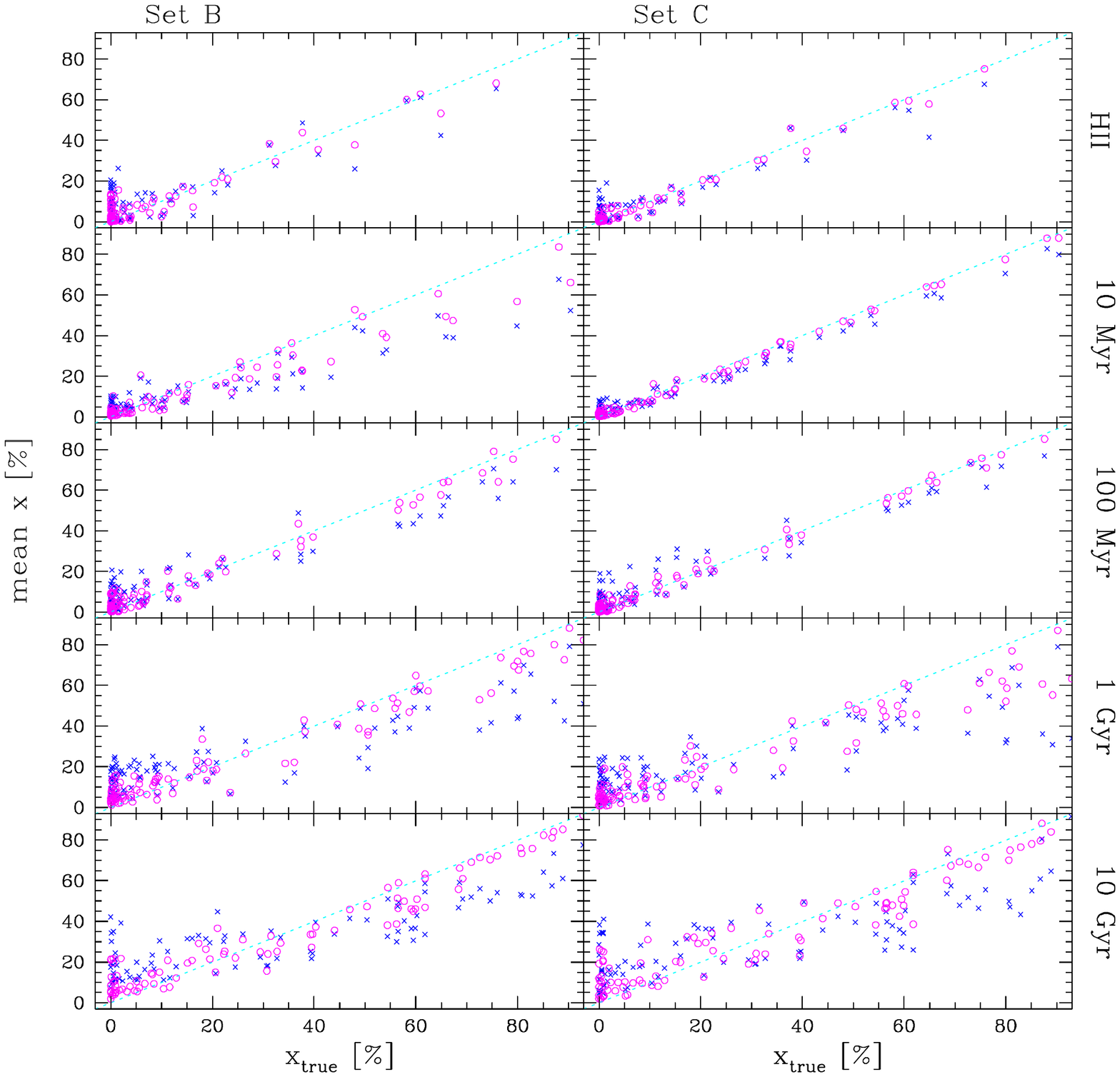}}
\caption{Input $\times$ output age binned ${\bf x}$ fractions for Set
B (left) and Set C (right) observables. Circles correspond to $S/N =
100$, and crosses to $S/N = 30$.}
\label{fig:Age_grouped_input_X_ouput_B_and_C}
\end{figure*}

This conclusion also holds for Sets B and C, as illustrated in
Fig.~\ref{fig:Age_grouped_input_X_ouput_B_and_C}. Naturally, the
restricted input data in these sets translates into a loss of
information, and the agreement is not as good as for Set A.  Note that
Set C does a much better job in recovering populations of 100 Myr or
less than Set B, a further example of how useful colors are in the
synthesis. This happens because the color information (present in Set
C but absent in B) impose stronger constraints on young populations
than in older systems.

%***FIG***FIG***FIG***FIG***FIG***FIG***FIG***FIG***FIG***FIG***FIG
%***FIG***FIG***FIG***FIG***FIG***FIG***FIG***FIG***FIG***FIG***FIG

The reliability of age-grouped proportions contrasts with the badly
constrained and biased results for the individual components of ${\bf
x}$, and is one of the main results of our study. This result puts in
perspective all previous interpretations of the synthesis with this
base. This fundamental limitation of the base was in fact previously
known, but was never studied in detail. Schmitt \etal (1999), for
instance, preferred to summarize the results of their synthesis study
in terms of age grouped proportions for populations younger than 1
Gyr, in consonance with the results above. Still, in that work we kept
a distinction between the different $Z$'s in the 10 Gyr age-group. At
this level of detail the population fractions are very uncertain. In
fact, since Set C observables and $S/N \sim 30$--60 were used in that
study, Fig.~\ref{fig:Age_grouped_input_X_ouput_B_and_C} indicates that
it would be safer to group their 1 and 10 Gyr proportions. Yet, the
proportions for the younger ($\le 100$ Myr) populations found by
Schmitt \etal (1999), and which constituted the focus of that paper,
are trustworthy (Fig.~\ref{fig:Age_grouped_input_X_ouput_B_and_C}).

\subsubsection{The age $\times$ metallicity degeneracy}

\label{sec:Age_X_Metallicity_degeneracy}

A close inspection of the $S/N = 10$ and 30 results in
Fig.~\ref{fig:Age_and_Z_grouped_input_X_ouput} reveals a compensation
effect between the 1 and 10 Gyr age bins, albeit at a much smaller
level than for the individual components
(Fig.~\ref{fig:individual_input_X_ouput}). This effect is much more
pronounced in Sets B and C, as seen in
Fig.~\ref{fig:Age_grouped_input_X_ouput_B_and_C}.

The confusion between the 1 and 10 Gyr populations is related to the
age-$Z$ degeneracy, which sets in precisely in this age range. This
well known effect (O'Connell 1986, 1994; Worthey 1994) is also present
in Bica's base and affects the synthesis results. To visualize and
quantify this effect we have computed the average log $Z$ and log age
for both the input and synthetic ${\bf x}$ for our test galaxies.
Given that ${\bf x}$ is a light fraction at 5870 \AA, these averages
ultimately represent flux-weighted ages and metallicities, which serve
our current illustrative purposes, despite their ambiguous physical
significance.

\begin{figure*}
%\resizebox{\textwidth}{!}{\includegraphics{BWFig_Age_X_Z_degeneracy.eps}}
\resizebox{\textwidth}{!}{\includegraphics{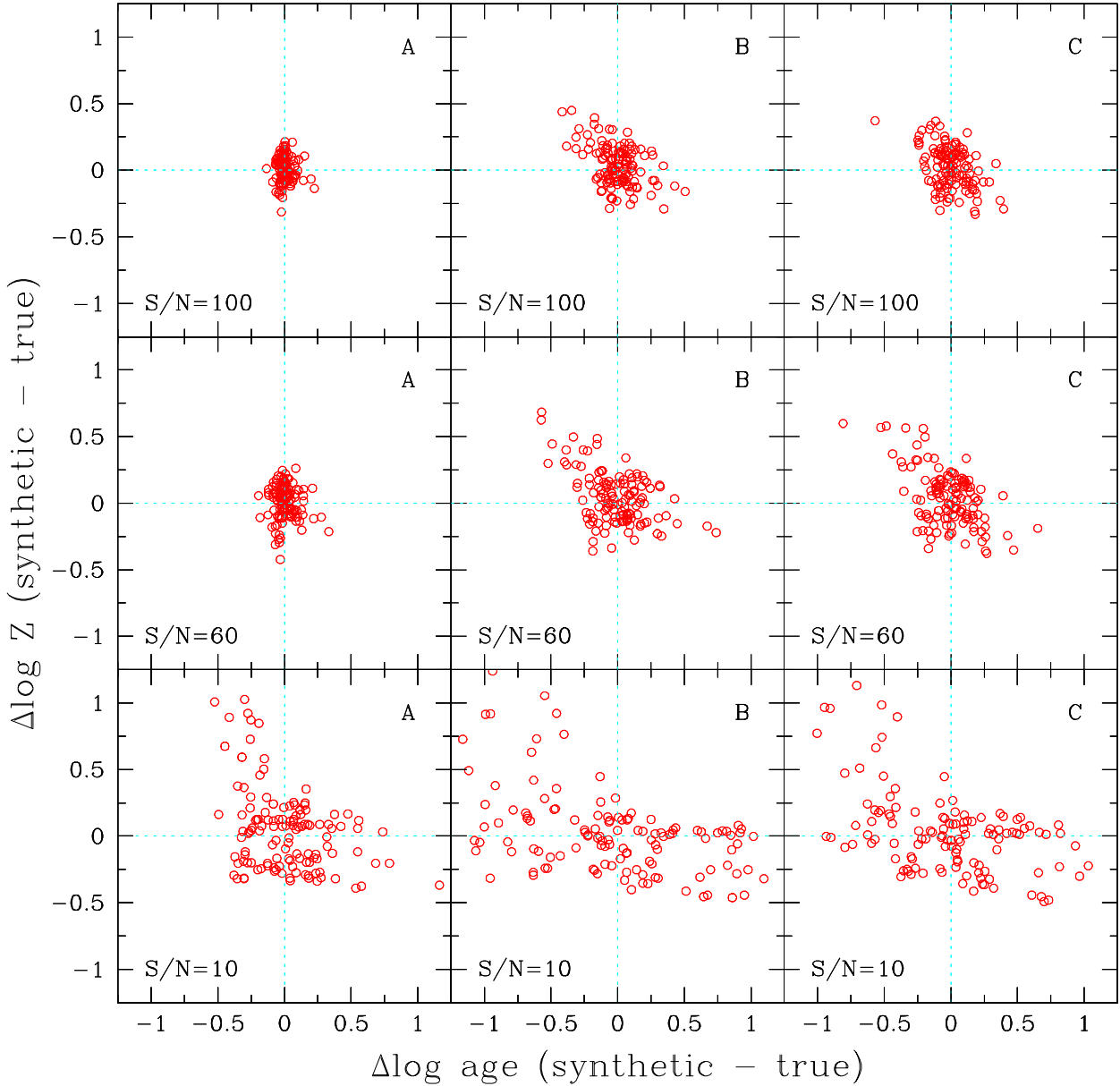}}
\caption{The age-metallicity degeneracy. The x-axis maps the
difference between the output and input log age, with the
corresponding difference in $\log Z$ plotted along the y-axis. Circles
correspond to age and $Z$ computed from the mean (${\bf x},A_V$),
whereas dots correspond to the best model sampled. All 126 test
galaxies are plotted. The different panels correspond to combinations
of the data quality ($S/N$, labeled in the bottom-left corners) and
the set of observables (top-right) used in the synthesis.
}
\label{fig:Age_X_Z_degeneracy}
\end{figure*}

The `output - input' $\log {\rm age}$ and $\log Z$ residuals are
plotted against each other in Fig.~\ref{fig:Age_X_Z_degeneracy}. In
these diagrams $\Delta \log {\rm age} > 0$ (overestimated age) implies
redder colors and stronger metal lines, which tend to be
counter-balanced by the bluer colors and weaker lines resulting from
$\Delta \log Z < 0$ (underestimated $Z$). The age-$Z$ degeneracy is
most obvious in the results for Sets B and C, for which the biases in
age and $Z$ persist even for $S/N =100$.  For Set A, on the other
hand, the residuals are small: $\lapprox$ 0.2 dex for $S/N$ $\gapprox$
60, which shows that the age-$Z$ degeneracy can be broken, at coarse
the level of age and $Z$ resolution offered by the base, with enough
information and good spectra. Note that this remark applies to the
global averages defined above, not the detailed component by component
ages and $Z$'s, which we concluded to be highly uncertain
(Fig.~\ref{fig:individual_input_X_ouput}).

%***FIG***FIG***FIG***FIG***FIG***FIG***FIG***FIG***FIG***FIG***FIG
%***FIG***FIG***FIG***FIG***FIG***FIG***FIG***FIG***FIG***FIG***FIG

\subsection{Metallicity grouped proportions}

%***FIG***FIG***FIG***FIG***FIG***FIG***FIG***FIG***FIG***FIG***FIG
\begin{figure}
%\resizebox{\textwidth}{!}{\includegraphics{BWFig_inXout_Zgrouped_ave_E.eps}}
%\resizebox{\textwidth}{!}{\includegraphics{fig12.eps}}
\resizebox{15cm}{!}{\includegraphics{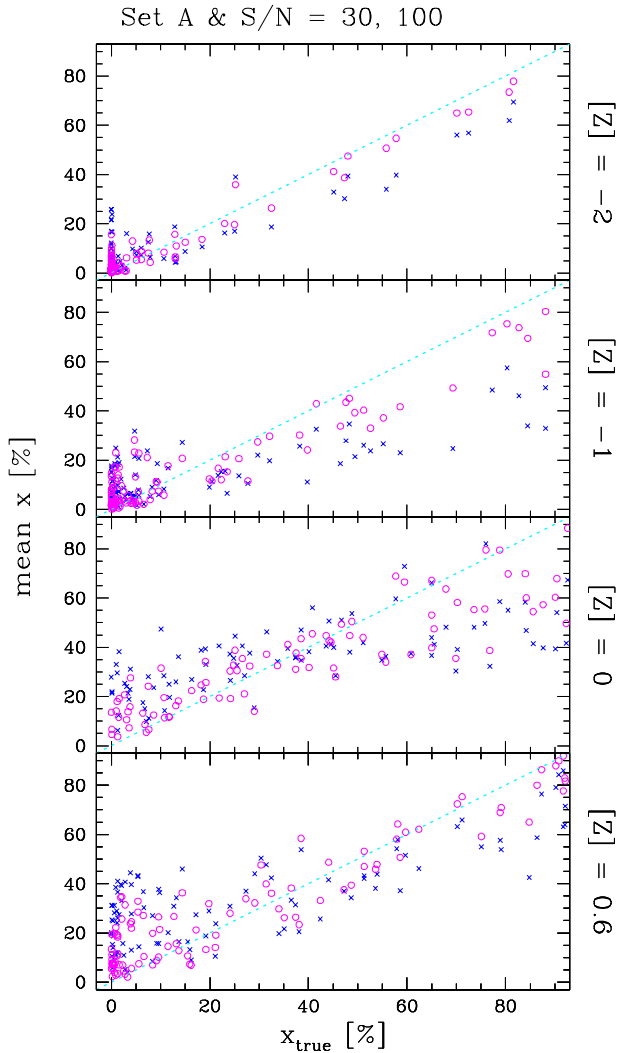}}
\caption{Metallicity binned light fractions in 100 input test galaxies
versus the corresponding synthetic means. Circles correspond to $S/N =
100$, and crosses to $S/N = 30$, both for Set A observables.  The
`resolution power' of the base is much better for ages
(Fig.~\ref{fig:Age_and_Z_grouped_input_X_ouput}) than for
metallicities.}
\label{fig:Z_grouped_input_X_ouput}
\end{figure}
%***FIG***FIG***FIG***FIG***FIG***FIG***FIG***FIG***FIG***FIG***FIG

The results for $Z$ binned proportions, plotted in
Fig.~\ref{fig:Z_grouped_input_X_ouput}, are not nearly as good as
those for the age binned groups. With the exception of the $Z =
10^{-2} Z_\odot$ bin, which is actually represented by just one
element (Table 1), the scatter in the input $\times$ output $Z$-binned
proportions is so large for all other $Z$'s that one is forced to
conclude that no accurate description of the chemical history of
galaxies can be afforded with this spectral base. Given this limited
$Z$ `resolution power', one might consider removing intermediate
components such as 2, 3 and 6.  However, as we concluded that only the
age distribution can be assessed with the base, there is no obvious
advantage in this further reduction. This situation can be improved by
providing {\it a priori} constraints on the occupancy of the age-$Z$
plane, like those imposed by B88 based on chemical evolution
arguments, since these effectively reduce the dimensionality of the
base to typically 8 components, thus alleviating the confusion between
components and producing better focused results. The validity of this
prior input external to the synthesis process, nonetheless, has to be
evaluated in an object by object basis. In this study we followed S91
in not imposing any such extra information, as this approach
encompasses a larger class of evolutionary scenarios and thus
contemplates more possible applications.

\subsection{Best $\times$ Mean Parameters}

\label{sec:Best_X_Mean}

Throughout these experiments, we have consistently used the {\it
mean} parameters as an estimate of the result of the synthesis.  This
convention was deliberately chosen to follow the more widely employed
version of EPS with this base and thus to allow a reassessment of
previous results.  Furthermore, the mean is a convenient and
reasonable way to summarize the results insofar as it represents a
`center of mass' of acceptable solutions. The severe biases
identified in the mean synthetic population vector, however, raise the
question of whether alternative approaches should be pursued instead.

By definition, minimization oriented techniques would do a better job
at recovering the input parameters. Indeed, we verified that the best
model population vector sampled in the Markov chain is much closer to
${\bf x}_{true}$ than its expected value, an agreement which can only
be improved given that the Metropolis sampler is not an optimal
minimization procedure.  Though the simulations confirmed our
expectation that for small errors the mean and best solutions should
both converge to the true parameters, it was surprising to realize how
easily $\ov{\bf x}$ moves away from ${\bf x}_{true}$, and consequently
also from ${\bf x}_{best}$.  Very small uncertainties in the
observables, at the levels of $S/N > 300$, are enough to set powerful
compensation effects into operation (e.g.,
Figs.~\ref{fig:Errors_in_observables} and
\ref{fig:Correlation_Matrix}), shifting the components of $\ov{\bf x}$
away from their true values due to the asymmetric redistribution of
the probability in ${\bf x}$-space.

Minimization procedures can in principle overcome this bias, as
illustrated by the success of the S91 and Pelat (1997) tests as well
as by our (not optimized) results for ${\bf x}_{best}$.  A further
bias in mean solutions is that they always produce $\ov{x_i} >0$.
Real galaxies, even if synthesizable by the base, are likely to have
observables outside synthetic domain due to measurement errors. As
shown by Pelat (1997, 1998), the best model in this case lies at the
surface of the synthetic domain, where very often several of the ${\bf
x}$-components are 0.  In these aspects, minimization algorithms are
more suitable than the more traditional sampling approach.  However,
the high susceptibility of the synthesis to the data quality indicates
that minor perturbations in the input observables, within their
errors, may seriously affect the estimation of the best model
parameters. The stability of the best solution is thus likely to be
more fragile than that of the mean solution, which takes into account
the effects of the errors in the observables in the trade-off between
an optimal and a statistically acceptable fit of the data. We have in
fact detected this effect in a series of simulations, but a more
careful study, with specialized minimization techniques is required to
quantify it properly. It would be particularly interesting to carry
out this test with the method developed by Pelat (1997) applied to the
base used in this work.  His Monte Carlo tests with a $n_\star = 10$
stellar base and $n_W = 19$ equivalent widths produced essentially
unbiased population fractions even for $S/N$ as low as 10. Extending
his formalism to synthesize galaxy colors would also be desirable.

In any case, because of the nature of the EPS problem and the
structure of Bica's base, relatively large discrepancies in the
parameters do not necessarily translate into large differences in the
synthesized quantities.  In this respect, the difference between mean
and best models is largely academic, as the mean solution already does
an excellent job in fitting the observables
(Fig.~\ref{fig:input_X_output_observables}).

\section{Summary}

\label{sec:conclusions}

In this paper we have revisited the method of Empirical Population
Synthesis with the aims of improving it both at the formal and
computational levels, and exploring its efficacy as a tool to probe
the population mixture of galaxies.  Our results can be divided into
three parts.

{\bf (1)} A simple probabilistic formulation of the problem was
presented, which puts this method, traditionally employed in a less
formal way, onto a mathematical footing. It was shown how former
applications of the method fit into the probabilistic formulation,
thus providing some {\it a posteriori} justification for previous
results.

{\bf (2)} An importance sampling scheme, based on the Metropolis
algorithm, was developed and tested. It provides a more efficient and
smooth mapping of the probability distribution of the parameters than
is possible with the commonly used uniform grid sampling.

{\bf (3)} In the third part of the paper, we applied the formulation
and sampling method to a series of test galaxies constructed out the
$n_\star = 12$ star-clusters base of S91. The tests explored the
ability to reconstruct the model parameters, which carry a record of
the star formation and chemical evolution in galaxies, in the presence
of {\it (i)} observational errors and {\it (ii)} limited data. This
study centered on the comparison of input parameters with the {\it
mean} synthetic solution ($\ov{\bf x},\ov{A_V}$), as this is the most
common form of EPS in the literature with this base, and a systematic
evaluation of the consistency of this method has not been carried out
before. The main results of this study can be summarized as follows:

\begin{itemize}

\item[]{\bf (a)} The allowance for errors in the observables sparkle
linear dependences within the base elements. This induces systematic
{\it biases} in $\ov{\bf x}$, redistributing the probability in
components with a large fractional contribution to the integrated
spectrum among similar components. This happens preferentially for
components of same age. Though this was a predictable result, it was
surprising to realize how powerful this effect is. For any realistic
$S/N$, the bias is such that in most cases one cannot trust the
estimates of all 12 {\it individual} components of $\ov{\bf x}$ to any
useful level of confidence.

\item[]{\bf (b)} Reducing the number of observational constraints, due
to incomplete coverage of the 3300--9000 \AA\ spectral interval spanned
by the base, has the same overall effect as reducing the data quality,
that is, to broaden, shift and skew the probability distribution of
the $x_i$'s. We find that synthesizing colors as well as equivalent
widths yields substantially better results than synthesizing $W$'s
only. The need to account for the extinction as an extra parameter is
largely compensated by the better focused ${\bf x}$ obtained by
considering the color information, particularly for populations of 100
Myr and younger. Unlike for the individual population fractions, we do
not detect systematic biases in the estimates of $A_V$.

\item[]{\bf (c)} Despite the difficulty to retrieve accurately the
detailed population vector, the observables are very precisely
reproduced by the mean solution.  The mapping between the parameters
and observables spaces is thus highly `degenerate'.  The nature of
this `non-uniqueness' is not algebraic, as it happens also when the
number of free parameters exceeds the numbers of observables and in
undetermined cases where the solution is known to be unique. Instead,
it is related to the possibility of mimicking to a statistically
indistinguishable level the effects of certain base elements by linear
combinations of other elements. This happens even for tiny
observational errors, inducing a `spill over' of the likelihood among
base components, carrying the mean solution along.

\item[]{\bf (d)} The structure of the base reflects the fact that
evolution is the main source of variance in stellar populations.
Accordingly, we found that grouping the $\ov{\bf x}$ components in
their five age bins produces much more reliable results, and we
recommend this procedure when evaluating the detailed population
vector obtained in previously published results with this base. Some
confusion between the 1 and 10 Gyr populations also takes place due to
the age-$Z$ degeneracy, but this can be minimized with an adequate
spectral coverage (measuring all observables in the base) and $S/N
\simgreat 30$. Grouping the results for the four different $Z$'s in
the base is also recommended, but caution must be exerted when
interpreting the results since these are much less precise than the
age-binned proportions.

\end{itemize}

The difficulties to retrieve accurately the true stellar population
parameters stem from two sources: {\it (i)} The convention to
represent the results of the synthesis by the mean solution, and {\it
(ii)} limitations imposed noise-induced linear dependences in the
base. Minimization techniques may be instrumental in overcoming the
biases in the mean population vector, which ultimately limit the
resolution of the 12-D base to its 5 age bins. Within our
probabilistic formulation, other priors can be investigated. Maximum
Entropy, for example, is a prior that has been applied successfully to
inverse problems formally similar to population synthesis, like image
restoration. However, more than improvements on the method, which are
possible and desirable, further work on the base itself will certainly
be needed.  Indeed, the limited age and $Z$ `resolution power' of the
base employed here lies at the heart of all difficulties identified in
this paper. Including other spectral indices, modeling the full
spectrum and incorporating other relevant information such as mass to
light ratios in the synthesis process are possibles ways to progress in
this direction.

\section*{ACKNOWLEDGEMENTS}

We thank Catherine Boisson, Monique Joly, Jihane Moultaka and Didier
Pelat for stimulating discussions on the virtues and drawbacks of EPS
techniques. RCF thanks the hospitality of Johns Hopkins University,
where this work was finalized. Support for this work was provided by
the National Science Foundation through grant \# GF-1001-99 from the
Association of Universities for Research in Astronomy, Inc., under NSF
cooperative agreement AST-9613615.  JRSL acknowledges an MSc
fellowship awarded by CAPES. Partial support from CNPq, FAPESP, PRONEX
and FUNPESQUISA-UFSC is also acknowledged.

\end{document}